\documentclass{amsart}
\usepackage{fullpage,graphicx,subfigure,mathpazo,color}
\usepackage{palatino}  %字体
\usepackage{amsmath,amscd,tikz}
\usepackage[normalem]{ulem}
\usepackage{amsmath}
\usepackage{amssymb}
\usepackage{setspace}%使用间距宏包
\usepackage{CJKutf8}%最后要删除
\usepackage{cite}
\usepackage{verbatim}%注释
\usepackage[colorli nks,
 urlcolor=blue,
 linkcolor=blue,
 anchorcolor=blue,
 citecolor=blue]{hyperref}
\newcommand{\ii}{\mathrm{i}}
\newcommand{\ee}{\mathrm{e}}
\newcommand{\dd}{\mathrm{d}}
\newcommand{\T}{\mathrm{T}}

\newcommand{\jy}[1]{{\color{black} #1}}
\newcommand{\djy}[1]{{\color{black} #1}}

\makeatletter
\newtheorem{theorem}{Theorem}

\newtheorem{proposition}{Proposition}

\newtheorem{remark}{Remark}
\newtheorem{rhp}{Riemann-Hilbert problem}

\usepackage{graphicx}      % insert graphic
\usepackage{titlesec}
\graphicspath{{figure/}}
\titleformat{\section}{\centering\LARGE\bfseries}{\thesection}{1em}{}
\titleformat{\subsection}{\Large\bfseries}{\thesubsection}{1em}{}
\newcommand{\setParDis}{\setlength {\parskip} {0.5cm} }
\newcommand{\setParDef}{\setlength {\parskip} {0pt} }
\begin{document}

\title{Kadomtsev-Petviashvili equation: one-constraint method and lump pattern}

\author{Jie-Yang Dong}
\address{School of Mathematics, South China University of Technology, Guangzhou, China, 510641}
\author{Liming Ling}
\address{School of Mathematics, South China University of Technology, Guangzhou, China, 510641}
\author{Xiaoen Zhang}
\address{School of Mathematics, South China University of Technology, Guangzhou, China, 510641}
\email{Corresponding author:zhangxiaoen@scut.edu.cn}

\begin{abstract}
The Kadomtsev-Petviashvili reduction method is a crucial method to derive the solitonic solutions of $(1+1)$ dimensional integrable system from high dimensional system. In this work, we explore to use the solutions of lower dimensional system to construct the solutions in the high dimensional one with the Darboux transformation. Especially, we utilize this method to disclose the relationship between the rogue wave and lump solutions. Under one-constraint method, the asymptotic analysis to the lump pattern of Kadomtsev-Petviashvili equation is given.

\end{abstract}

\date{\today}

\maketitle

\section{Introduction}\label{sec1}
As one of the most important integrable $(2+1)$ dimensional equations, the Kadomtsev-Petviashvili (KP) equation has a wide applications, it can be used to
%is a well-known integrable model and can be utilized to
describe the capillary gravitational waves on a liquid surface and magneto-acoustic waves in plasma \cite{kumar2013bifurcations,saha2015bifurcation,wang2012dynamics}. In general, the KP equation is written in the form
\begin{equation}\label{KP}
\left(4 u_{t}-12 u u_{x}-u_{x x x}\right)_{x}+3 u_{y y}=0,
\end{equation}
where $u=u(x, y, t)$ denotes a scalar function with respect to the variables $x, y$ and $t$.
%Compared to the classical Korteweg-de Vries (KdV) equation and nonlinear Schr\"{o}dinger(NLS) equation, the KP equation is a high-dimensional system, which will present much more attractive properties in some aspects.
Similar to the $(1+1)$ integrable equation, the KP equation also has Lax pair, bilinear form and the symmetry constraint \cite{lou1997infinitely}. Due to its integrability, there are lots of research on it, including the soliton solution \cite{chakravarty2009soliton,ablowitz1991solitons}, the lump solution \cite{manakov1977twodimensional,johnson1978solution,satsuma1979two,Krichever78} and the quasi-periodic solution \cite{biondini2008kadomtsev}, the corresponding methods refer to the Darboux transformation \cite{terng2000backlund, ablowitz2000novel}, the bilinear method \cite{kodama2004young}, the inverse scattering method \cite{ablowitz1991solitons} and the algebraic geometric method \cite{Dubrovin80}. Among these interesting solutions, the lump solution is a special one, which is a type of the localized traveling waves in both $x$ and $y$ directions\cite{He2017proA,He2020CNSNS,He2021PhysicaD,He2019PLA}.

In \cite{dubard2011multi}, the authors constructed the high order lump solutions for KP equation and high order rogue wave solutions for nonlinear Schr\"{o}dinger(NLS) equation respectively, and the result indicates that the high order rogue waves of NLS equation is similar to the high order lump solutions of KP equation. This incredible phenomena can be explained by the theory of ``k-constraint". Actually, this theory is very important to understand the KP equation and it can derive the KP hierarchy to the nonlinear system for finite number of dynamical coordinates \cite{cheng1992constraints,cheng1991constraint,konopelchenko19911+,konopelchenko1991akns,He2014JGP,He2006Sigma,He2003KP}. Especially, under the framework of Sato theory \cite{sato1983nonlinear} with one-constraint, the AKNS hierarchy can be derived through the KP hierarchy. As a result, the solution $u(x,y,t)$ of KP equation can be represented as a product of $\psi(x,y,t)\psi^*(x,y,t)$, in which, the function $\psi(x,y,t)$ satisfies the second and the third flow of AKNS system. In \cite{konopelchenko1991akns}, the authors gave a detailed description about the relation between the KP equation and the AKNS flows, which presents a good confirmation about the property between the rogue wave and lump solution again. Moreover, this method had been applied to other aspects, such as in \cite{Cao1999KP}, the authors constructed the quasiperiodic solution to KP equation with the nonlinearized Zakharov-Shabat eigenvalue problem. Apart from the KP equation, this one-constraint method can also be used to other $(2+1)$ dimensional integrable equations \cite{Geng2001}.

In \cite{kodama2004young}, Kodama gave the interaction patterns of $N$-soliton with the Young diagrams and gave a classification to the $N$ soliton solutions. When $t$ is large, the high order soliton or multi-solitons will be split into single soliton. However, the rogue wave is different from the soliton on the asymptotic analysis. As to the rogue wave in $(1+1)$ dimensional system, when $t$ is large, the rogue wave will go back to the constant background, thus its asymptotic behavior with respect to the variable $t$ is meaningless. Whereas, the geometry structure and the location character about the rogue waves are more diverse than the soliton, which is found to be determined by some free parameters. Very recently, in \cite{yang2021rogue,yang2021universal}, the authors analyzed the asymptotics of rogue wave with respect to these parameters and gave a general conclusion about the location distribution. Inspired by this theory, we would like to analyze the lump patten of KP equation as well as its asymptotics. With the one-constraint method, the solution of KP equation is connected with the AKNS system, thus we try to analyze this asymptotics with the aid of the Darboux transformation of high dimensional AKNS system. As we know that the Darboux transformation of the AKNS system has a perfect architecture and theoretical framework, which has its merit in contrast to the Darboux transformation of the Lax pair of KP equation. Under the framework of this kind of Darboux transformation, the maximum of solution can be analyzed easily, which has not been realized by the original Darboux transformation of KP equation.

%As we know that the KP equation admits the Darboux transformation, so the exact solutions can be constructed by this transformation in a determinant form. But the maximum value of the solution can not been proved by the determinant formula directly.  For the Darboux transformation of ANKS system with the version of loop group action, we just obtain a partial of solutions under the constraints, but the benefit is that we can prove the maximum for the solutions and find a way to construct the solutions with the largest amplitude.

This paper is organized as follows. In Section \ref{sec2}, we give a brief introduction about the KP hierarchy under one-constraint condition and convert the Lax pair of KP equation to a high dimensional AKNS system. With the one-constraint method, we construct the Darboux transformation to derive the solution formula in Section \ref{sec3}. By choosing a special spectral parameter under the plane wave background, we obtain the high order lump solution and rewrite it to a $\tau$ function obtained by Ohta and Yang \cite{ohta2012general}. In Section \ref{sec4}, we analyze the lump patterns from several aspects, one is the asymptotics with respect to the time variable $t$ and the other one is the asymptotics with one parameter $a_{2m+1}$, additionally, by the dynamics of solution at $t=0$, we display a classification for lump pattern. We also give an auxiliary material in the appendix.

\section{The KP hierarchy under one-constraint}\label{sec2}
In this section, we intend to present some introduction about the KP equation and the one-constraint method. It is well known that the KP hierarchy can be constructed with the Sato theory and the corresponding introduction had been reported in many previous works. The Sato theory is originated from the Kyoto school, which can be regarded as one of the most famous theories in the integrable system. In view of this point, the KP hierarchy is the fundamental one, and many other integrable equations can be derived from the KP hierarchies. Such as, the KdV equation, the Boussinesq equation, the NLS equation and so on. To analyze the KP equation, we first give a brief review about how the Lax pair of KP equation can be derived from the microdifferential operator \cite{cheng1992constraints,date1983transformation,ohta1988elementary,sato1983nonlinear}. Suppose the operator $L$ as
\begin{equation}\label{L}
L=\partial+u_{2} \partial^{-1}+u_{3} \partial^{-2}+u_{4} \partial^{-3}+\cdots,\quad \partial=\partial/ \partial_{x},
\end{equation}
where $u_{i}$, $i=2,3,\cdots$ are functions with respect to the variables  $\mathbf{t}=(t_{1},t_{2},t_{3},\cdots)$.
Consider a system of linear equations about the eigenfunction $\psi$,
\begin{subequations}\label{psi-kp}
	\begin{align}
	L \psi&=\lambda \psi, \\
	\psi_{t_{n}}&=B_{n} \psi,\,\,\, B_n=(L^n)_+,
	\end{align}
\end{subequations}
where $\lambda$ is the spectral parameter, the subscript $_+$ denotes the positive part of the microdifferential operator. Through the compatibility conditions for the system \eqref{psi-kp}, we can derive the Lax equation as
\begin{equation}\label{lax equation}
L_{t_{n}}=\left[B_{n}, L\right],
\end{equation}
where $B_n$ denotes the differential part of $L^{n}$ and can be uniquely determined by the coordinates $u_{2},u_{3},\cdots$ and their $x$ derivatives. With a simple calculation, the first few operators $B_{n}$ can be given as
\begin{equation}\label{eq:B}
B_{1}=\partial,\quad B_{2}=\partial^{2}+2 u_{2},\quad B_{3}=\partial^{3}+3 u_{2} \partial+3 u_{3}+3 u_{2, x}.\\
\end{equation}

Based on the Lax equation \eqref{lax equation}, the KP hierarchy can be determined recursively. For later analysis, we give a generalized Leibniz rule,
\begin{equation}\label{Leibniz}
\partial^{n} f(x) =\sum\limits_{j=1}^{\infty} \frac{n(n-1)(n-2) \cdots(n-j+1)}{j !} \frac{d^{j} f}{d x^{j}} \partial^{n-j},
\end{equation}
where $n$ is an integer. By choosing $n=2$ and $n=3$ in Eq.(\ref{lax equation}), we can obtain the following two equations:
\begin{subequations}\label{u22}
\begin{align}\label{u22}
	u_{2, t_{2}}&=u_{2, x x}+2u_{3, x}, \\\label{u23}
	u_{3, t_{2}}&=u_{3, x x}+2 u_{4, x}+2 u_{2} u_{2, x},
\end{align}
\end{subequations}
and
\begin{subequations}
	\begin{align}\label{u32}
	u_{2, t_{3}}&=u_{2, x x x}+3 u_{3, x x}+3 u_{4, x}+6 u_{2} u_{2, x}, \\
	u_{3, t_{3}}&=u_{3, x x x}+3 u_{4, x x}+3 u_{5, x}+6\left(u_{2} u_{3}\right)_{x},
	\end{align}
\end{subequations}
where the subscript $_x$ represents the derivative with respect to $x$. Obviously, the functions $u_3$ and $u_4$ can be expressed as a function of $u_2$ via the Eq.(\ref{u22}) and Eq.(\ref{u23}),
\begin{equation}\label{u3}
\begin{split}
u_{3}&=\frac{1}{2} \partial^{-1}(u_{2, t_{2}}- u_{2,xx}),\\
u_{4}&=\frac{1}{4} \partial^{-2}\left(u_{2, t_{2} t_{2}}-u_{2, t_{2} xx}\right)-\frac{1}{4}\left(u_{2,t_{2}} -u_{2, x x}\right)-\partial^{-1}\left(u_{2} u_{2, x}\right).
\end{split}
\end{equation}
Next, substitute Eq.(\ref{u3}) to Eq.(\ref{u32}) and set $u_{2}=u$, $t_{1}=x$, $t_{2}=-\ii y$, $t_{3}=t$, the KP equation (\ref{KP}) can be derived.

The Lax pair and its adjoint Lax pair for the KP equation are given by:
\begin{subequations}
	\begin{align}\label{linear11}
	(-\ii)\psi_{y}&=\psi_{x x}+2 u \psi, \\\label{linear12}
	\psi_{t}&=\psi_{x x x}+3 u \psi_{x}+\frac{3}{2}\left(u_{x} -\ii \partial_{x}^{-1} u_{y}\right) \psi,
	\end{align}
\end{subequations}
and
\begin{subequations}
	\begin{align}\label{linear21}
	(-\ii)\psi^{*}_{y}&=-\psi^{*}_{x x}-2 u \psi^{*}, \\\label{linear22}
	\psi^{*}_{t}&=\psi^{*}_{x x x}+3 u \psi^{*}_{x}+\frac{3}{2}\left(u_{x}-\ii \partial_{x}^{-1} u_{y}\right) \psi^{*},
	\end{align}
\end{subequations}
where the superscript $^*$ represents the complex conjugate.
If we identify the conserved covariant $u$ with the covariant generator $\psi \psi^{*}$,
\begin{equation}\label{qr}
u=\psi \psi^{*}=:q r,
\end{equation}
where $q \equiv \psi$ and $r \equiv \psi^{*}$. Then Eq.(\ref{linear11}) and Eq.(\ref{linear21}) can be reduced into
\begin{equation}\label{simplify2}
\begin{split}
\ii q_{y}&= -q_{x x}-2 q^{2} r, \\
\ii r_{y}&= r_{x x}+2  q r^{2}.
\end{split}
\end{equation}
From Eq.\eqref{simplify2}, we can easily get the identity $\ii (qr)_y=(qr_x-q_x r)_x$.
Together with Eq.(\ref{linear12}) and Eq.(\ref{linear22}), we have
\begin{equation}\label{simplify3}
\begin{split}
q_{t}&=q_{x x x}+6 q_{x} q r,\\
r_{t}&=r_{x x x}+6q r r_{x}.
\end{split}
\end{equation}
It can be seen that Eq.(\ref{simplify2}) and Eq.(\ref{simplify3}) are the second and third flow of the AKNS hierarchy \cite{cheng1992constraints}. Thus if $q$ and $r$ satisfy Eq.(\ref{simplify2}) and Eq.(\ref{simplify3}), then the potential $u$ will satisfy KP equation Eq.(\ref{KP}). Therefore, the solutions of KP equation can be derived from AKNS system \cite{cheng1992constraints,konopelchenko1991akns}.

Now we give a simple introduction about the one-constraint of KP equation. In general, the operator of $k$-constraint is set as $ L^{k}=B_{k}+q \partial^{-1} r $ \cite{cheng1992constraints}. When $k=1$, the so called one-constraint operator is changed into
\begin{equation} \label{one-constraint}
L=\partial+q \partial^{-1} r.
\end{equation}
\begin{comment}
According to Eq.(\ref{L})(\ref{Leibniz}), we can express uniquely the coordinates $u_{2}$, $u_{3}$, $\cdots$ with $q$, $r$ and their $x$ derivatives.
\begin{equation}\label{simplify1}
u_{2}=q r, \quad u_{3}=-q r_{x},  \quad u_{i}=(-1)^{i}q\frac{d^{i-2}r}{dx^{i-2}}
\end{equation}
Considering the coupling system
\begin{equation}\label{Coupling system}
\begin{split}
q_{t_{n}}&=B_{n}q\\
r_{t_{n}}&=-B_{n}^{*}r
\end{split}
\end{equation}
With the constraint (\ref{one-constraint}) and set $t=2,3$, we can reduce (\ref{Coupling system}) into (\ref{simplify2}) and (\ref{simplify3})
respectively, which explain the constraint $u=qr$ with Sato theory. Specifically speaking, the constraint (\ref{qr}) is consistent with one-constraint.
%We can see that Eq.(\ref{simplify2}) and Eq.(\ref{simplify3}) are the first two system of the AKNS hierarchy.\upcite{cheng1992constraints}.    We can know the fact that if $q$ and $r$ satisfy the first two linear system Eq.(\ref{simplify2})(\ref{simplify3}) then the potential $u$ satisfies KP Eq.(\ref{KP}). Therefore, by carrying out the first two AKNS system successively, we can obtain a special lump solution  of the KP equation.\upcite{cheng1992constraints}\upcite{konopelchenko1991akns}
\end{comment}
With this operator, consider the following coupled eigenvalue problems:
\begin{subequations}
	\begin{align} \label{17a}
L \psi& =\psi_{x}+q \varphi=\lambda \psi, \\\label{psi}
\varphi_{x}&=r \psi,
\end{align}
\end{subequations}
and
\begin{equation}\label{eq18}
\begin{split}
\psi_{t_{n}}&=B_{n}\psi ,\\
\varphi_{t_{n}}&=A_{n}\psi,
\end{split}
\end{equation}
where  $\varphi$ is a new “eigenfunction” defined by (\ref{psi}) and $A_{n}$ is the polynomial in $\partial$ satisfying
\begin{equation}
\partial A_{n}=r B_{n}-\left(B_{n}^{*} r\right).
\end{equation}
Denote
\begin{equation}
\pmb{\Phi}(\lambda;x,\mathbf{t}) =\begin{bmatrix} \varphi(-2\ii\lambda;x,\mathbf{t}), -\ii \psi(-2\ii\lambda;x,\mathbf{t}) \end{bmatrix}^{\T}\ee^{\ii\lambda x},
\end{equation}
then the equations (\ref{17a}) and (\ref{eq18}) can be rewritten as
\begin{equation}\label{lax pair uv}
\pmb{\Phi}_{x}=\textbf{U}(\lambda; x, \mathbf{t})\pmb{\Phi},\quad \pmb{\Phi}_{t_{n}}=\mathbf{V}_{n}(\lambda; x, \mathbf{t})\pmb{\Phi},
\end{equation}
where
\begin{equation}
\mathbf{U}(\lambda; x, \mathbf{t})= \mathrm{i}\left(\lambda \sigma_{3}+\mathbf{Q}\right),\,\,\,\, \sigma_{3}=\begin{pmatrix}
1 & 0 \\
0 & -1
\end{pmatrix}, \quad \mathbf{Q}=\begin{pmatrix}
0 & r \\
q & 0
\end{pmatrix},\,\,\,\,\, \mathbf{V}_{n}(\lambda; x, \mathbf{t})=\begin{pmatrix}A&B\\
C&D
\end{pmatrix},
\end{equation}
$A, B,C,D$ are the polynomials with $\lambda$.
The system \eqref{lax pair uv} is the AKNS hierarchy with $\rm{su}(2)$ symmetry. Furthermore, we can get the second and third flow with the compatibility condition of Eq.\eqref{lax pair uv} by setting $n=2, 3$ and $t_{2}=-\ii y$, $t_{3}=t$, then the corresponding Lax pair of Eq.\eqref{simplify2} and Eq.\eqref{simplify3} can be given as
%Utilizing $\textbf{U}$ and set $n=1,2,3$, we can get
\begin{equation}\label{lax pair}
  \pmb{\Phi}_{x}=\textbf{U}\pmb{\Phi},   \,\,\,\,
  \pmb{\Phi}_{y}=\textbf{V}\pmb{\Phi}, \,\,\,\,
  \pmb{\Phi}_{t}=\textbf{W}\pmb{\Phi},
\end{equation}
where
\begin{equation}
\begin{split}
\mathbf{V}&= 2\left(\mathrm{i} \lambda^{2} \sigma_{3}+\mathrm{i} \lambda \mathbf{Q}-\frac{\mathrm{i}}{2} \sigma_{3} \mathbf{Q}^{2}+\frac{\sigma_{3}}{2} \mathbf{Q}_{x}\right), \\
\mathbf{W}&= -4 \mathrm{i} \lambda^{3} \sigma_{3}-4 \mathrm{i} \lambda^{2} \mathbf{Q}+2 \lambda\left(\mathrm{i} \sigma_{3} \mathbf{Q}^{2}-\sigma_{3} \mathbf{Q}_{x}\right)+\mathrm{i}  \mathbf{Q}_{x x}
+2 i \mathbf{Q}^{3}+ \mathbf{Q}_{x}  \mathbf{Q}- \mathbf{Q}  \mathbf{Q}_{x}.
\end{split}
\end{equation}
As we all know that the traditional Lax pair of KP equation is Eq.\eqref{linear11}, but in this paper, we will use the Lax pair of high dimensional AKNS system \eqref{lax pair}, which is more useful for analyzing the properties of solution further.

From the Lax pair Eq.\eqref{lax pair}, we can obtain the following equations through the compatibility conditions:
\begin{subequations}
	\begin{align}\label{condition1}
\mathbf{Z}^{[1]}&=\mathbf{U}_{y}-\mathbf{V}_{x}+[\mathbf{U},\mathbf{V}]=\mathbf{0},\\\label{condition2}
\mathbf{Z}^{[2]}&=\mathbf{U}_{t}-\mathbf{W}_{x}+[\mathbf{U},\mathbf{W}]=\mathbf{0},\\\label{condition3}
\mathbf{Z}^{[3]}&=\mathbf{V}_{t}-\mathbf{W}_{y}+[\mathbf{V}, \mathbf{W}]=\mathbf{0}.
\end{align}
\end{subequations}
With a simple calculation, Eq.(\ref{condition1}) and (\ref{condition2}) can be written as
\begin{equation}\label{Z1Z2}
\left\{\begin{split}
{\mathbf{Z}^{[1]}_{12} = \ii r_{y}-r_{x x}-2 q r^{2}=0 }, \\
{\mathbf{Z}^{[1]}_{21} =\ii q_{y}+q_{x x}+2 q^{2} r=0 },
\end{split} \quad \left\{\begin{split}
\mathbf{Z}^{[2]}_{12} = r_{t}-r_{x x x} -6 r q r_{x}=0, \\
\mathbf{Z}^{[2]}_{21} =q_{t}-q_{x x x}-6 r q q_{x} =0.
\end{split}\right.\right.
\end{equation}
Moreover, $\mathbf{Z}^{[3]}$ can be given as
\begin{equation}\label{Z3}
\begin{split}
	\mathbf{Z}^{[3]}&=\begin{pmatrix}
	\mathbf{Z}^{[3]}_{11} & \mathbf{Z}^{[3]}_{12}\\
	\mathbf{Z}^{[3]}_{21}  & \mathbf{Z}^{[3]}_{22}
	\end{pmatrix}
	,\quad \mathbf{Z}^{[3]}_{22}=-\mathbf{Z}^{[3]}_{11},\\
	\mathbf{Z}^{[3]}_{11}&=-2\left( q\mathbf{Z}^{[1]}_{12}+r\mathbf{Z}^{[1]}_{21}\right) \lambda-\ii \mathbf{Z}^{[1]}_{12}+\ii r_{x}\mathbf{Z}^{[1]}_{21}+\ii q \left(\mathbf{Z}^{[1]}_{12}\right)_{x}-\ii r \mathbf{Z}^{[1]}_{21}-\ii q \mathbf{Z}^{[2]}_{12}-\ii r\mathbf{Z}^{[2]}_{21},\\
	\mathbf{Z}^{[3]}_{12}&=4\mathbf{Z}^{[1]}_{12}\lambda ^{2}-2\ii \left[\left(\mathbf{Z}^{[1]}_{12}\right)_{x}-\mathbf{Z}^{[2]}_{12}\right]\lambda-4qr\mathbf{Z}^{[1]}_{12}-2r^{2}\mathbf{Z}^{[1]}_{21}-\left( \mathbf{Z}^{[1]}_{12}\right) _{xx}+\left(\mathbf{Z}^{[2]}_{12}\right)_{x},\\
	\mathbf{Z}^{[3]}_{21}&=4\mathbf{Z}^{[1]}_{21}\lambda ^{2}+2 \ii \left[\left(\mathbf{Z}^{[1]}_{21}\right)_{x}+\mathbf{Z}^{[2]}_{21}\right]\lambda-2q^{2}\mathbf{Z}^{[1]}_{12}-4qr\mathbf{Z}^{[1]}_{21}-\left( \mathbf{Z}^{[1]}_{21}\right) _{xx}-\left( \mathbf{Z}^{[2]}_{21}\right) _{x}.
	\end{split}
\end{equation}
%where the subscript $x$ represents $x$ derivatives.
From Eq.(\ref{Z3}), we can see that if $q, r$ satisfy Eq.(\ref{Z1Z2}), they must satisfy Eq.(\ref{condition3}). Consequently, we only need to consider Eq.(\ref{Z1Z2}), then the solutions of KP equation can be given.
%And we can see that  Eq.(\ref{Z1Z2}) is is consistent with (\ref{simplify2}) and (\ref{simplify3}), which explain the constraint $u=qr$ with Sato theory.
%Specifically speaking, the constraint (\ref{qr}) is consistent with one-constraint.

\section{The one-constraint method to the KP equation}\label{sec3}
Last section, under the framework of Sato theory, we give the one-constraint of KP equation in a high dimensional AKNS system. During the calculation, we first give a constraint condition $u=qr$, where $u$ is the solution of KP equation and $q$, $r$ can be derived from the AKNS system. Compared to the original Lax pair Eq.\eqref{linear11}, the study to the new Lax pair has much more advantages. Such as, to derive the lump solution, if we use the original Lax pair Eq.\eqref{linear11}, we should construct the binary Darboux transformation, but if we use the Lax pair of AKNS system, we only need to modify the well-known Darboux matrix of AKNS system, which seems more simple. More importantly, we can establish the Riemann-Hilbert problem with the Darboux matrix of AKNS system and continue to study the asymptotics, which is hard with the original Lax pair Eq.\eqref{linear11} to our knowledge. In\cite{Zhou1994HD,Zhou1996DS}, the authors studied the soliton solutions on high dimensional AKNS system through the Darboux transformation, including the N-wave equation, Davey–Stewartson(DS) equation. Inspired by this idea, in this section, we will construct the lump solution to KP equation with the Lax pair\eqref{lax pair}.

The Darboux transformation for the AKNS system with ${\rm su}(2)$ symmetry is given by
\begin{equation}
\mathbf{T}_{1}(\lambda ; x,y, t)=\mathbb{I}-\frac{\lambda_{1}-\lambda_{1}^{*}}{\lambda-\lambda_{1}^{*}} \mathbf{P}_{1}(x,y, t), \quad  \mathbf{P}_{1}(x,y, t)=\frac{\phi_{1} \phi_{1}^{\dagger} }{\phi_{1}^{\dagger} \phi_{1}},
\end{equation}
where
\begin{equation}\label{alphabeta}
\phi_{1}:=[\phi_{1,1},
\phi_{1,2}]^{\T}
=c_1(x,y,t)\pmb{\Phi}\left(\lambda_{1} ; x, y,t\right) \pmb{\Phi}^{-1}\left(\lambda_{1} ; 0, 0,0\right)(1,-\mathrm{i})^{T}=c_1(x,y,t)\pmb{\Phi}\left(\lambda_{1} ; x, y,t\right)(\alpha_{1},\beta_{1})^{T}\end{equation}
and $c_1(x,y,t)$ is an arbitrary function on $x$, $y$ and $t$.
It can be verified that the Darboux matrix admits the following symmetry
\begin{equation}
\mathbf{T}_{1}(\lambda; x, y, t) \mathbf{T}_{1}^{\dagger}\left(\lambda^{*};  x, y, t\right) =\mathbb{I}
\end{equation}
according to the symmetry of the Lax pair \eqref{lax pair}: $\mathbf{U}^{\dag}(\lambda^*;x,y,t)=-\mathbf{U}(\lambda^*;x,y,t)$, $\mathbf{V}^{\dag}(\lambda^*;x,y,t)=-\mathbf{V}(\lambda^*;x,y,t)$ and $\mathbf{W}^{\dag}(\lambda^*;x,y,t)=-\mathbf{W}(\lambda^*;x,y,t)$.
Then the B\"{a}cklund transformation about the potential functions can be given by
\begin{equation}
\mathbf{Q}^{[1]}=\mathbf{Q}-\left(  \lambda_{1}-\lambda_{1}^{*}\right) \left[\mathbf{P}_{1}, \sigma_{3}\right],
\end{equation}
which indicates the following identity:
\begin{equation}\label{EQ36}
q^{[1]}(x,y,t)=q(x,y,t)-2(\lambda_{1}-\lambda_{1}^{*})\frac{\phi_{1,2}\left(\lambda_{1}; x, y,t\right) \phi_{1,1}^{*}\left(\lambda_{1},x,y,t\right)}{\phi_{1}^{\dagger}\left( \lambda_{1};x,y,t\right) \phi_{1}\left(\lambda_{1};x,y,t\right)}.
\end{equation}
From this expression, we can get the maximal peak by the mean inequality, which is shown in the following proposition.
\begin{proposition}\label{prop1}
	The maximal peak about the new solution $|q^{[1]}(x,y,t)|$ is given by
	\begin{equation}
	\max_{x,y,t}(|q^{[1]}(x,y,t)|)=\max_{x,y,t}(|q(x,y,t)|)+2\,|{\rm Im}(\lambda_1)|,
	\end{equation}
in particular, by set the special condition $\phi_{1}=\Phi\left(\lambda_{1} ; x, y,t\right) \Phi^{-1}\left(\lambda_{1} ; 0, 0,0\right)(1,-\mathrm{i})^{T}$, the maximum point is located at the origin $(x,y,t)=(0,0,0)$.
\end{proposition}
\begin{remark}
Similar to the properties of rogue waves, the maximum modulus of $q^{[N]}$ is also $1+2N|{\rm Im}(\lambda_1)|$, which is given by Proposition \ref{prop1}.  Meanwhile, this kind of Darboux transformation provides a way to construct the high order lump solution with the maximal peak.
\end{remark}

To construct the lump solution, we set the plane wave seed solution of $q$ and $r$ in Eq.(\ref{lax pair}) as:
\begin{equation}\label{solution}
\begin{array}{l}
q=b\ee^{\ii \theta},\quad
r=b\ee^{- \ii \theta},\quad
\theta=c x+\left(2 b^{2}-c^{2}\right) y+\left(6 b^{2} c-c^{3}\right)t,
\end{array}
\end{equation}
where $b,c$ are arbitrary real number. Substituting the above special seed solution \eqref{solution} to the corresponding Lax equation Eq.\eqref{lax pair}, then we have the fundamental solution
\begin{equation}\label{Eq38}
\begin{split}
\mathbf{\Phi}(\lambda;x,y,t)=\ee^{-\frac{\mathrm{i}}{2}\theta\sigma_3}\mathbf{E}\ee^{\mathrm{i}\xi\omega\sigma_{3}},\,\,\,\, \mathbf{E}\equiv \begin{pmatrix}
1 & 1 \\
\frac{b}{\frac{1}{2}(c+2 \lambda)+\xi} & \frac{b}{\frac{1}{2}(c+2 \lambda)-\xi}
\end{pmatrix} ,\quad
\xi=\sqrt{\left(\lambda+\frac{c}{2}\right)^2+b^2},\\ \omega=x+(2 \lambda-c) y+2\left(b^{2}-\frac{1}{2} c^{2}+c \lambda-2 \lambda^{2}\right) t+a(\lambda),
\end{split}
\end{equation}
where $a(\lambda)$ is independent with $x$, $y$ and $t$. Without loss of generality, we can set $b=1, c=0$, other types of choice are equivalent to each other because of the Galilean symmetry:
\begin{equation}
\tilde{q}(x, y, t)=\ee^{-\frac{1}{9}\ii\left(3x\epsilon-y\epsilon^2-\frac{1}{3}t\epsilon^3\right)}q\left(x-\frac{2}{3}y\epsilon-\frac{1}{3}t\epsilon^2, y+t\epsilon, t\right).
\end{equation}

Moreover, by iterating the above Darboux transformation step by step, then the $N$-fold Darboux transformation can be represented as a compact form. And the multi-solitonic solution can be derived with the $N$-fold Darboux transformation.
With the theory of Darboux transformation for the AKNS system, we give the $N$-fold Darboux transformation for the system (\ref{lax pair}) in Theorem \ref{thm:DT}.
	\begin{theorem}\label{thm:DT}
	Suppose there is a bounded smooth solution $q(x, y,t) \in \mathbf{L}^{\infty}\left(\mathbb{R}^{3}\right) \cap \mathbf{C}^{\infty}\left(\mathbb{R}^{3}\right)$. And
	the matrix solution $\pmb{\Phi}(\lambda;x,y,t)$  is analytic in the whole complex plane $\mathbb{C}$, then the $N$-fold Darboux transformation can be represented as
 	\begin{equation}
	\mathbf{T}_{N}(\lambda ; x, y,t)=\mathbb{I}+\mathbf{Y}_{N} \mathbf{M}^{-1} \mathbf{D}^{-1}\mathbf{Y}_{N}^{\dagger},
	\end{equation}
	where
	\begin{equation}\label{DT-def}
	\begin{split}
		\mathbf{D}&={\rm diag}\left(\lambda-\lambda_1^*,\lambda-\lambda_2^*,\cdots,\lambda-\lambda_N^*\right),\\
		\mathbf{Y}_N&=\left[\phi_1,\phi_2,\cdots,\phi_N\right],\,\,\,\,\, \mathbf{M}=\left(\frac{\phi_i^{\dag}\phi_j}{\lambda_i^*-\lambda_j}\right)_{1\leq i,j\leq N},\\
\phi_{i}&=[\phi_{i,1},
\phi_{i,2}]^{\T}
=c_i(x,y,t)\pmb{\Phi}\left(\lambda_{i} ; x, y,t\right) \pmb{\Phi}^{-1}\left(\lambda_{i} ; 0, 0,0\right)(1,-\mathrm{i})^{T},
	\end{split}
	\end{equation}
then the new potential can be given with the following B\"{a}cklund transformation
	\begin{equation}\label{B transformation}
	{q}^{[N]}={q}+2 \mathbf{Y}_{N, 2} \mathbf{M}^{-1} \mathbf{Y}_{N, 1}^{\dagger},
	\end{equation}
	where the subscript $\mathbf{Y}_{N, i}$ denotes the $i$-th
	row vector of $\mathbf{Y}_{N}$, $i=1,2$.
\end{theorem}

\begin{remark}
In view of the Darboux transformation, by choosing different spectral parameter $\lambda_i$, we can get different types of solutions. If $\lambda_i=\alpha\ii$ with $0<\alpha<1$, we can get the breathers which is localized in $y$-direction and periodic in $x$-direction. If $\lambda_i=\alpha\ii$ with $\alpha>1$, we can get the breathers which is localized in $x$-direction and periodic in $y$-direction. If $\lambda_i+\lambda_i^*\neq0$, the other types of breathers can be obtained. High order ones and their mixtures can also be obtained by performing the above transformation. \end{remark}

Our idea is planning to analyze the asymptotics by using the $\tau$ function given by Ohta and Yang. To realize it, we should derive the $\tau$ function from the Darboux transformation. Through the Eq. (\ref{EQ36}), (\ref{solution}), the solution $q^{[1]}$ can be rewritten as
\begin{equation}
q^{[1]}= \ee^{2\mathrm{i} y}+2\left(\lambda_{1}^{*}-\lambda_{1}\right) \frac{\phi_{1,2} \phi_{1,1}^{*}}{\phi_{1}^{\dagger} \phi_{1}}=\frac{\frac{\phi_{1}^{\dagger} \phi_{1} \ee^{2 \i y}}{-2\mathrm{i}\left(\lambda_{1}^{*}-\lambda_{1}\right)}+\mathrm{i}\phi_{1,2} \phi_{1,1}^{*}}{\frac{\phi_{1}^{\dagger} \phi_{1}}{-2\mathrm{i}\left(\lambda_{1}^{*}-\lambda_{1}\right)}}.
\end{equation}
By the property of determinant, the solution of Eq.\eqref{B transformation} can be represented in the following form:
\begin{equation}\label{mij}
\jy{q^{[N]}=\frac{\underset{1\leqslant i,j\leqslant N}{\det}\left( m_{ij}^{(1)}\right) }{\underset{1\leqslant i,j\leqslant N}{\det}\left( m_{ij}^{(0)}\right) }\ee^{2 \ii y} },\,\,\,\, m_{ij}^{(1)}=\frac{\phi_{i}^{\dagger} \phi_{j} \ee^{  2 \mathrm{i}  y}}{-2 \mathrm{i}\left(\lambda_{i}^{*}-\lambda_{j}\right)}+\mathrm{i} \phi_{j,2} \phi_{i,1}^{*},\,\,\, m_{ij}^{(0)}=\frac{\phi_{i}^{\dagger} \phi_{j}}{-2 \mathrm{i}\left(\lambda_{i}^{*}-\lambda_{j}\right)}.
\end{equation}
Moreover, from the definition of $\phi_{i}$ in Eq.\eqref{DT-def}, the component $m^{(0)}_{ij}$ and $m^{(1)}_{ij}$ can be given in a quadric form
\begin{equation}\label{mij1}
m_{ij}^{(0)}=c_i^*c_j\left(\alpha_{i}^{*} \ee^{-\mathrm{i} \xi_{i}^{*} \omega_{i}^{*}}, \beta_{i}^{*} \ee^{\mathrm{i} \xi_{i}^{*} \omega_{i}^{*}}\right)\begin{pmatrix}
\frac{1}{-\mathrm{i}(\lambda_{i}^{*}+\xi_{i}^{*})+\mathrm{i}\left(\lambda_{j}+\xi_{j}\right)} & \frac{1}{-\mathrm{i}(\lambda_{i}^{*}+\xi_{i}^{*})+\mathrm{i}\left(\lambda_{j}-\xi_{j}\right)} \\
\frac{1}{-\mathrm{i}\left(\lambda_{i}^{*}-\xi_{i}^{*}\right)+\mathrm{i}(\lambda_{j}+\xi_{j})} & \frac{1}{-\mathrm{i}\left(\lambda_{i}^{*}-\xi_{i}^{*}\right)+\mathrm{i}\left(\lambda_{j}-\xi_{j}\right)}
\end{pmatrix}
\begin{pmatrix}
\alpha_{j} \ee^{\mathrm{i} \xi_{j} \omega_{j}} \\
\beta_{j} \ee^{-\mathrm{i} \xi_{j} \omega_{j}}
\end{pmatrix},
\end{equation}
\begin{equation}\label{mij0}
m_{ij}^{(1)}=c_i^*c_j\left(\alpha_{i}^{*} \ee^{-\mathrm{i} \xi_{i}^{*} \omega_{i}^{*}}, \beta_{i}^{*} \ee^{\mathrm{i} \xi_{i}^{*} \omega_{i}^{*}}\right)\begin{pmatrix}
\frac{\frac{\lambda_{i}^{*}+\xi_{i}^{*}}{\lambda_{j}+\xi_{j}}}{-\mathrm{i}(\lambda_{i}^{*}+\xi_{i}^{*})+\mathrm{i}\left(\lambda_{j}+\xi_{j}\right)} & \frac{\frac{\lambda_{i}^{*}+\xi_{i}^{*}}{\lambda_{j}-\xi_{j}}}{-\mathrm{i}(\lambda_{i}^{*}+\xi_{i}^{*})+\mathrm{i}\left(\lambda_{j}-\xi_{j}\right)} \\
\frac{\frac{\lambda_{i}^{*}-\xi_{i}^{*}}{\lambda_{j}+\xi_{j}}}{-\mathrm{i}(\lambda_{i}^{*}-\xi_{i}^{*})+\mathrm{i}\left(\lambda_{j}+\xi_{j}\right)} & \frac{\frac{\lambda_{i}^{*}-\xi_{i}^{*}}{\lambda_{j}-\xi_{j}}}{-\mathrm{i}\left(\lambda_{i}^{*}-\xi_{i}^{*}\right)+\mathrm{i}\left(\lambda_{j}-\xi_{j}\right)}
\end{pmatrix}
\begin{pmatrix}
\alpha_{j} \ee^{\mathrm{i} \xi_{j} \omega_{j}} \\
\beta_{j} \ee^{-\mathrm{i} \xi_{j} \omega_{j}}
\end{pmatrix}.
\end{equation}
For simplicity, we introduce some new notations
\begin{equation}\label{Eq46}
p_{i}^{*}=-\mathrm{i}\left(\lambda_{i}^{*}+\xi_{i}^{*}\right) \quad q_{j}=\mathrm{i}\left(\lambda_{j}+\xi_{j}\right).
\end{equation}
According to the definition of $\alpha_{i},\beta_{j}$ in Eq.\eqref{alphabeta} and the relationship $\xi ^{2}=1+\lambda ^{2}$, we have
\begin{equation}
\alpha_{i}^{*}=\ii (p_{i}^{*}+1), \quad \beta_{i}^{*}=-\ii \left( \frac{1}{p_{i}^{*}}+1\right),\quad \alpha_{j}=-\ii(q_{j}+1), \quad \beta_{j}=\ii \left(\frac{1}{q_{j}}+1\right).
\end{equation}
By choosing the proper parameters $c_i$, we know that the exponent term $\mathrm{i}\xi_{j} \omega_{j}-\mathrm{i} \xi_{i}^{*} \omega_{i}^{*}$ can be changed into another equivalent form, that is
\begin{equation}
\begin{split}\label{EQ45}
\mathrm{i}\xi_{j} \omega_{j}-\mathrm{i} \xi_{i}^{*} \omega_{i}^{*}
\sim&\left[\left(\ii \lambda_{j}+\mathrm{i} \xi_{j}\right) \left( x+a\left( \lambda_j\right) \right) -\left( \mathrm{i}\left( \lambda_{j}+\xi_{j}\right) \right) ^{2}\mathrm{i}y+\left( \left( \mathrm{i}\left( \lambda_{j}+\xi_{j}\right) \right) ^{3}+3\mathrm{i}(\lambda_{j}+\xi_{j})\right) t\right]\\
+&\left [-\mathrm{i}\left( \lambda_{i}^{*}+\xi_{i}^{*}\right) \left( x+a^{*}\left( \lambda_i\right) \right) +\left( \left( -\mathrm{i}\left( \lambda_{i}^{*}+\xi_{i}^{*}\right) \right) ^{2}\mathrm{i}y+\left( -\mathrm{i}\left( \lambda_{i}^{*}+\xi_{i}^{*}\right) \right) ^{3}-3\mathrm{i}\left( \lambda_{i}^{*}+\xi_{i}^{*}\right) \right) t\right]\\
=&\left[q_{j}x-q_{j}^{2}\mathrm{i}y+\left( q_{j}^{3}+3q_{j}\right) t+\sum\limits_{k=0}^{\infty}\hat{a}_{k}\left( \ln q_{j}\right) ^{k}\right]+\left[p_{i}^{*}x+p_{i}^{*2}\mathrm{i}y+\left( p_{i}^{*3}+3p_{i}^{*}\right) t+\sum\limits_{k=0}^{\infty}\hat{a}_{k}^{*}\left( \ln p_{i}^{*}\right) ^{k}\right],
\end{split}
\end{equation}
where $a(\lambda_j)=q_j^{-1}\sum\limits_{k=0}^{\infty}\hat{a}_{k}(\ln q_{j})^{k}$.

Furthermore, set the factor involving the exponent term $\ee^{\mathrm{i}\xi_{j} \omega_{j}-\mathrm{i} \xi_{i}^{*} \omega_{i}^{*}}$ in Eq.(\ref{mij1}) and Eq.(\ref{mij0}) as \djy{$\hat{B}_{ij}^{(0)}$} and $\hat{B}_{ij}^{(1)}$ respectively. With this definition, we introduce some new notations $\zeta_i^*, \eta_{j}$ as
\begin{equation}
\zeta_{i}^{*}=p_{i}^{*} x+p_{i}^{*2} \mathrm{i} y+\left(p_{i}^{*3}+3 p_{i}^{*}\right) t,\quad \eta_{j}=q_{j} x-q_{j}^{2} \mathrm{i} y+\left(q_{j}^{3}+3 q_{j}\right) t,
\end{equation}
then we have
\begin{equation}\label{eq:bij}
\begin{split}
	\hat{B}_{ij}^{(0)}(p_{i}^{*}, q_{j}) &= \frac{(p_{i}^{*}+1)(q_{j}+1)}{p_{i}^{*}+q_{j}} \exp\left(\zeta_{i}^{*}+\eta_{j}+\sum\limits_{k=0}^{\infty}\hat{a}_{k}^{*}(\ln p_{i}^{*})^{k}+\sum\limits_{k=0}^{\infty}\hat{a}_{k}(\ln q_{j})^{k}\right),\\
	\hat{B}_{ij}^{(1)}(p_{i}^{*}, q_{j}) &= \frac{(p_{i}^{*}+1)(q_{j}+1)}{p_{i}^{*}+q_{j}} \left(-\frac{p_{i}^{*}}{q_{j}}\right)\exp\left(\zeta_{i}^{*}+\eta_{j}+\sum\limits_{k=0}^{\infty}\hat{a}_{k}^{*}(\ln p_{i}^{*})^{k}+\sum\limits_{k=0}^{\infty}\hat{a}_{k}(\ln q_{j})^{k}\right).
	\end{split}
\end{equation}
Next, we make a minor revision to $\hat{B}_{ij}^{(0)}$ and $\hat{B}_{ij}^{(1)}$ with the purpose of not altering the value of $q^{[N]}$ in Eq.(\ref{mij}), that is
\begin{equation}
	B_{ij}^{(n)}(p_{i}^{*}, q_{j})=\frac{	\hat{B}_{ij}^{(n)}(p_{i}^{*}, q_{j})}{\exp\left( \sum\limits_{k=0}^{\infty}A^{+}_{2k}(\ln p_{i}^{*})^{2k}+\sum\limits_{l=0}^{\infty}A^{-}_{2l}(\ln q_{j})^{2l}\right) },\quad n=0,1,
\end{equation}
where $A^{+}_{k},A^{-}_{l}$ are given by the following expansion:
\begin{equation}
 \zeta_{i}^{*}+\eta_{j}+\sum\limits_{k=0}^{\infty}\hat{a}_{k}^{*}(\ln p_{i}^{*})^{k}+\sum\limits_{k=0}^{\infty}\hat{a}_{k}(\ln q_{j})^{k}=\sum\limits_{k=0}^{\infty}A^{+}_{k}(\ln p_{i}^{*})^{k}+\sum\limits_{l=0}^{\infty}A^{-}_{l}(\ln q_{j})^{l},
 \end{equation}
i.e.
\begin{equation}\label{EQ50}
	A^{+}_{k}=\frac{x+2^{k}\ii y+3^{k}t+3t}{k!}+\hat{a}_{k}^{*},\quad 	A^{-}_{l}=\frac{x-2^{l}\ii y+3^{l}t+3t}{l!}+\hat{a}_{l}.
\end{equation}
In Eq. \eqref{eq:bij},  the even power terms of $\ln q_j$ and $\ln p_i^*$ in exponential factor can be eliminated since the elements determinants are invariant under the transformation $\ln q_j\to-\ln q_j$ and $\ln p_i^*\to-\ln p_i^*$.
We can rewrite $B_{ij}^{(n)}(p_{i}^{*}, q_{j})$ as
\begin{equation}\label{EQ51}
	\frac{1}{1-\frac{(p_{i}^{*}-1)(q_{j}-1)}{( p_{i}^{*}+1)( q_{j}+1)}}(-1)^{n}\exp \left(  \sum\limits_{k=0}^{\infty}A^{+}_{2k+1}(\ln p_{i}^{*})^{2k+1}+n(\ln p_{i}^{*})+\sum\limits_{l=0}^{\infty}A^{-}_{2l+1}(\ln q_{j})^{2l+1}-n(\ln q_{j}) \right).
\end{equation}

Under the condition $\lambda_{i}=\lambda_{j}=-\ii$, we have $p_{i}^{*}=q_{j}=1, \xi_{i}=\xi_{j}=0.$  From the definition $m_{ij}^{(n)}(n=0,1)$ in Eq.\eqref{mij1} and Eq.\eqref{mij0}, we get a relation between $m_{ij}^{(n)}$ and $\tau$ function
\begin{equation}
\begin{split}
\frac{m_{ij}^{(n)}}{\exp\left( \sum\limits_{k=0}^{\infty}A^{+}_{2k}(\ln p_{i}^{*})^{2k}+\sum\limits_{l=0}^{\infty}A^{-}_{2l}(\ln q_{j})^{2l}\right) } &=B_{ij}^{(n)}(p_{i}^{*}, q_{j})-B_{ij}^{(n)}\left(p_{i}^{*}, \frac{1}{q_{j}}\right)-B_{ij}^{(n)}\left(\frac{1}{p_{i}^{*}}, q_{j}\right)+B_{ij}^{(n)}\left(\frac{1}{p_{i}^{*}}, \frac{1}{q_{j}}\right)\\
&= \djy{4\sum_{k,l=0}^{\infty} \tau_{2 k+1,2 l+1}^{(n)}(\ln p_{i}^{*})^{2 k+1}(\ln q_{j})^{2l+1},}
\end{split}
\end{equation}
where \djy{$B_{ij}^{(n)}=\sum\limits_{k, l=0}^{\infty} \tau_{k,l}^{(n)}(\ln p_{i}^{*})^{k}(\ln q_{j})^{l}$} is expanded at $\lambda_{i}=\lambda_{j}=-\ii$, and
\begin{equation}\label{tau1}
\djy{\tau_{k,l}^{(n)}=\left.\frac{1}{k!}(\partial_{\ln p_{i}^{*}})^{k}\frac{1}{l!}(\partial_{\ln q_{j}})^{l}B_{ij}^{(n)}\right|_{p_{i}^{*}=q_{j}=1},\quad n=0,1.}
\end{equation}
 %Similarly, $B_{ij}^{(1)}$ can also be given as $B_{ij}^{(1)}=\sum\limits_{m, n=0}^{\infty} \tau_{m,n}^{(1)}(\ln p_{i}^{*})^{m}(\ln q_{j})^{n}$ with
%\begin{equation}\label{tau0}
%\begin{split}
%\tau_{m,n}^{(1)}=&\djy{\frac{1}{B_{i,j}^{(1)}}}\left.\frac{1}{m!}(\partial_{\ln p_{i}^{*}})^{m}\frac{1}{n!}(\partial_{\ln q_{j}})^{n}B_{ij}^{(1)}\right|_{p_{i}^{*}=q_{j}=1},\\
%m_{ij}^{(1)}=&4\sum_{m,n=0}^{\infty} \tau_{2 m+1,2 n+1}^{(1)}(\ln p_{i}^{*})^{2 m+1}(\ln q_{j})^{2 n+1}.
%\end{split}
%\end{equation}

Correspondingly, the solution $q^{[N]}$ in Eq. \eqref{mij} can be converted into the following form:
\begin{equation}
\begin{split}
	q^{[N]}=\ee^{2\ii y}\frac{\det\left( \mathbf{P}\boldsymbol{\tau}^{(1)}\mathbf{Q}\right) }{\det\left( \mathbf{P}\boldsymbol{\tau}^{(0)}\mathbf{Q}\right)  },\quad    \boldsymbol{\tau}^{(n)}=\begin{bmatrix}
	\tau_{1,1}^{(n)} &\tau_{1,3}^{(n)}& \cdots & \tau_{1,2N-1}^{(n)} & \cdots \\
\tau_{3,1}^{(n)} &\tau_{3,3}^{(n)} & \cdots & \tau_{3,2N-1}^{(n)}&\cdots \\
	\vdots &\vdots & &\vdots\\\tau_{2N-1,1}^{(n)} & \tau_{2N-1,3}^{(n)} & \cdots &\tau_{2N-1,2N-1}^{(n)} & \cdots\\
\vdots & \vdots & &\vdots
	\end{bmatrix},\quad n=0,1  \\
	\mathbf{P}=\begin{bmatrix}
	\ln p_{1}^{*} & (\ln p_{1}^{*})^{3} & \cdots & \left(\ln p_{1}^{*}\right)^{2 N-1} & \cdots \\
	\ln p_{2}^{*} & \left(\ln p_{2}^{*}\right)^{3} & \cdots & \left(\ln p_{2}^{*}\right)^{2 N-1}&\cdots \\
	\vdots &\vdots & &\vdots\\
	\ln p_{N}^{*} & \left(\ln p_{N}^{*}\right)^{3} & \cdots & \left(\ln p_{N}^{*}\right)^{2 N-1} & \cdots
	\end{bmatrix} ,\quad \mathbf{Q}=\begin{bmatrix}
	\ln q_{1} & \ln q_{2} & \cdots & \ln q_{ N} \\
	(\ln q_{1})^{3} &(\ln q_{2})^{3} & \cdots & (\ln q_{N})^{3} \\
	\vdots &\vdots & &\vdots\\
	(\ln q_{1})^{2N-1} &(\ln q_{2})^{2N-1} & \cdots &\ln q_{N}^{2N-1}\\
	\vdots & \vdots & & \vdots
	\end{bmatrix},
\end{split}
\end{equation}
especially, when $p,q\rightarrow 1$, $\ln p^{*}, \ln q\rightarrow 0 $, the solution $q^{[N]}$ will be simplified into the following formula.
\begin{equation}\label{sigman}
	q^{[N]}=\ee^{2\ii y}\frac{\det\left((\mathbf{P})_{N,N}(\boldsymbol{\tau}^{(1)})_{N,N}(\mathbf{Q})_{N,N}\right) }{\det\left((\mathbf{P})_{N,N}(\boldsymbol{\tau}^{(0)})_{N,N}(\mathbf{Q})_{N,N}\right) }=\ee^{2 \ii y}\frac{\det (\boldsymbol{\tau}^{(1)})_{N,N}}{\det (\boldsymbol{\tau}^{(0)})_{N,N}}:=\ee^{       2\ii y}\frac{\sigma_{1}}{\sigma_{0}},\quad \sigma_{n}=\underset{1\leqslant i,j\leqslant N}{\det}\left(\tau_{2 i-1,2 j-1}^{(n)}\right),
\end{equation}
where $\tau_{2i-1,2j-1}^{(n)}$ is defined in Eq.(\ref{tau1}), which is a function with respect to $x, y, t$ and $\hat{a}_{2k+1}, (k=1,2,\cdots, N-1)$. The determinant formula \eqref{sigman} is consistent with the result derived by the Hirota bilinear method in \cite{ohta2012general}. To the best of our knowledge, the universality of lump solution between the Darboux transformation and $\tau$ function had not been discovered in the previous research.

In fact, there are several distinct formulas on the rogue waves in the previous literatures based on different expansions. We show that they are equivalent to each other.
%The function $B^{(n)}$ is expanded in $\ln q$ and $\ln p$.
In the reference \cite{guo2012nonlinear}, Guo,  one of the authors and Liu used the expansion in $\epsilon$ by $\lambda=-\ii(1+\epsilon^2)$. Then we have
\begin{equation}
\ln p=\ln\left(1+\epsilon^2-\epsilon \sqrt{2+\epsilon^2}\right)=-\int_0^\epsilon \frac{2}{\sqrt{2+s^2}}\dd s=\sum_{i=1}^{\infty} p^{[i]} \epsilon^{2i-1},
\end{equation}
where $p^{[i]}=-\sqrt{2}\binom{-\frac{1}{2}}{i}\frac{1}{2^i(2i+1)}$. Similarly, the function
$\ln q=\ln\left(1+\eta^2-\eta \sqrt{2+\eta^2}\right)$ has the same expansion as $\eta$. Then we can obtain the following determinant formula
\begin{equation}
q^{[N]}=\ee^{2\ii y}\frac{\hat{\sigma}_1}{\hat{\sigma}_0},\,\,\, \hat{\sigma}_n=\underset{1\leqslant i,j\leqslant N}{\det}\left(\hat{\tau}_{2 i-1,2 j-1}^{(n)}\right),
\end{equation}
where $$m^{(n)}=\sum_{i=1, j=1}^{+\infty,+\infty}\hat{\tau}_{2 i-1,2 j-1}^{(n)}\epsilon^{2i-1}\eta^{2j-1}.$$
By the Fa\'a di Bruno formula, the coefficients between different expansions method have the following relation:
\begin{equation}
\hat{\tau}_{2i-1,2j-1}^{(n)}=\sum_{g=1,h=1}^{i,j}B_{2i-1,2g-1}B_{2j-1,2h-1}\tau_{2g-1,2h-1}^{(n)},
\end{equation}
where
\begin{equation}
B_{2i-1,2g-1}=\sum_{{\sum\limits_{s=1}^{i}(2s-1)k_s}=2i-1}^{\sum\limits_{s=1}^{i}k_s=2g-1}\frac{(2g-1)!}{k_1!k_2!\cdots k_i!}(p^{[1]})^{k_1}(p^{[2]})^{k_2}\cdots (p^{[n]})^{k_i},
\end{equation}
which indicates that
\begin{multline}
\begin{bmatrix}
\hat{\tau}_{1,1}^{(n)}&\hat{\tau}_{1,3}^{(n)} &\cdots &\hat{\tau}_{1,2N-1}^{(n)} \\
\hat{\tau}_{3,1}^{(n)}&\hat{\tau}_{3,3}^{(n)} &\cdots &\hat{\tau}_{3,2N-1}^{(n)} \\
\vdots&\vdots &\ddots &\vdots \\
\hat{\tau}_{2N-1,1}^{(n)}& \hat{\tau}_{2N-1,3}^{(n)}&\cdots &\hat{\tau}_{2N-1,2N-1}^{(n)} \\
\end{bmatrix}=\begin{bmatrix}
B_{1,1}^{(n)}&0 &\cdots &0 \\
B_{3,1}^{(n)}&B_{3,3}^{(n)} &\cdots &0 \\
\vdots&\vdots &\ddots &\vdots \\
B_{2N-1,1}^{(n)}& B_{2N-1,3}^{(n)}&\cdots &B_{2N-1,2N-1}^{(n)} \\
\end{bmatrix}\\
\cdot  \begin{bmatrix}
{\tau}_{1,1}^{(n)}&{\tau}_{1,3}^{(n)} &\cdots &{\tau}_{1,2N-1}^{(n)} \\
{\tau}_{3,1}^{(n)}&{\tau}_{3,3}^{(n)} &\cdots &{\tau}_{3,2N-1}^{(n)} \\
\vdots&\vdots &\ddots &\vdots \\
{\tau}_{2N-1,1}^{(n)}& {\tau}_{2N-1,3}^{(n)}&\cdots &{\tau}_{2N-1,2N-1}^{(n)} \\
\end{bmatrix}\begin{bmatrix}
B_{1,1}^{(n)}&B_{3,1}^{(n)} &\cdots &B_{2N-1,1}^{(n)} \\
0 &B_{3,3}^{(n)} &\cdots &B_{2N-1,3}^{(n)} \\
\vdots&\vdots &\ddots &\vdots \\
0 & 0 &\cdots &B_{2N-1,2N-1}^{(n)} \\
\end{bmatrix}.
\end{multline}
Thus we have $q^{[N]}=\ee^{2\ii y}\frac{\sigma_1}{\sigma_0}=\ee^{2\ii y}\frac{\hat{\sigma}_1}{\hat{\sigma}_0}.$

Furthermore, when $p_1=p_2=\cdots=p_N=1$, the $N$-th order Darboux matrix will be changed into the following form:
\begin{equation*}
\begin{split}
\mathbf{T}_{N}(\lambda; x,y, t)&=\mathbb{I}+\mathbf{Y}_N\mathbf{M}^{-1}\mathbf{D}\mathbf{Y}_{N}^{\dagger}, \qquad \mathbf{M}=\mathbf{X}^{\dagger}\mathbf{S}\mathbf{X}\\
\end{split}
\end{equation*}
and
\begin{equation*}
\begin{split}
{\mathbf{Y}}_{N}&=\left[\mathbf{\Phi}_{1}^{[0]},\mathbf{\Phi}_{1}^{[1]}, \cdots, \mathbf{\Phi}_{1}^{[N-1]}\right],\\
{\mathbf{D}}&=\begin{bmatrix}\frac{1}{\lambda-\ii}&0&\cdots&0\\
\frac{1}{(\lambda-\ii)^2}&\frac{1}{\lambda-\ii}&\cdots&0\\
\vdots&\vdots&\ddots&\vdots\\
\frac{1}{\left(\lambda-\ii\right)^{N}}&\frac{1}{\left(\lambda-\ii \right)^{N-1}}&\cdots&\frac{1}{\lambda-\ii}
\end{bmatrix},\qquad
{\mathbf{X}}=\begin{bmatrix}\mathbf{\Phi}_{1}^{[0]}&\mathbf{\Phi}_{1}^{[1]}&\cdots&\mathbf{\Phi}_{1}^{[N-1]}\\
0&\mathbf{\Phi}_{1}^{[0]}&\cdots&\mathbf{\Phi}_{1}^{[N-2]}\\
\vdots&\vdots&\ddots&\vdots\\
0&0&\cdots&\mathbf{\Phi}_{1}^{[0]}
\end{bmatrix},\\
{\mathbf{S}}&=\begin{bmatrix}
\binom{0}{0}\frac{\mathbb{I}_2}{2\ii}&\binom{1}{0}\frac{\mathbb{I}_2}{\left(2\ii \right)^2}&\cdots&\binom{N-1}{0}\frac{\mathbb{I}_2}{\left(2\ii \right)^{N}}\\
\binom{1}{1}\frac{(-1)\mathbb{I}_2}{\left(2\ii\right)^2}&\binom{2}{1}\frac{(-1)\mathbb{I}_2}{\left(2\ii\right)^3}&\cdots&\binom{N}{1}\frac{(-1)\mathbb{I}_2}{\left(2\ii \right)^{N+1}}\\
\vdots&\vdots&\ddots&\vdots\\
\binom{N-1}{N-1}\frac{(-1)^{N-1}\mathbb{I}_2}{\left(2\ii \right)^{N}}&\binom{N}{N-1}\frac{(-1)^{N-1}\mathbb{I}_2}{\left(2\ii \right)^{N+1}}&\cdots&\binom{2N-2}{N-1}\frac{(-1)^{N-1}\mathbb{I}_2}{\left(2\ii\right)^{2N-1}}
\end{bmatrix},
\end{split}
\end{equation*}
with $\pmb{\Phi}_1^{[k]}=\frac{1}{k!}\left(\frac{\rm d}{{\rm d}\lambda}\right)^{k}\pmb{\Phi}_1|_{\lambda=-\ii}$ and $$\pmb{\Phi}_1=\ee^{-\frac{\ii}{2}\theta\sigma_3}\left\{\frac{(\ii\lambda+1)\sin(\xi\omega)}{\xi}\begin{bmatrix}
1\\
\ii \\
\end{bmatrix}+\cos(\xi\omega)\begin{bmatrix}
1\\
-\ii \\
\end{bmatrix}\right\}.$$  To construct the high order lump solution in a compact form, we need to modify the expansion as shown in the previous section.
Unlike the previous binary Darboux transformation, we give the Darboux transformation via the Lax pair of high dimensional AKNS system, which can be used to construct the corresponding Riemann-Hilbert problem with the theory in \cite{Deniz2019,ling2021CNLS}. Through the $N$-th order Darboux matrix, we can define the following sectional analytic matrix
\begin{equation}
\mathbf{M}^{[N]}(\lambda; x, y, t):=\left\{\begin{split}&\mathbf{M}_{+}^{[N]}(\lambda; x, y, t)=\left(\frac{\lambda+\ii}{\lambda-\ii}\right)^{-N/2}{\mathbf{T}}_{N}(\lambda; x, y, t),\\
&\mathbf{M}_{-}^{[N]}(\lambda; x, y, t)={\mathbf{T}}_{N}(\lambda; x, y, t)\ee^{-\frac{\ii}{2}\theta\sigma_3}\mathbf{E}\ee^{\ii\xi\omega\sigma_3}\mathbf{E}^{-1}{\mathbf{T}}^{-1}_{N}(\lambda; 0, 0, 0)\mathbf{E}\ee^{-\ii\xi\omega\sigma_3}\mathbf{E}^{-1}\ee^{\frac{\ii}{2}\theta\sigma_3}.
\end{split}\right.
\end{equation}
Especially, if we choose $a(\lambda)=0$, then
\begin{equation}\label{DT-decom}
\left(\frac{\lambda+\ii}{\lambda-\ii}\right)^{-N/2}{\mathbf{T}}_{N}(\lambda; 0, 0, 0)=\mathbf{Q}_{c}\left(\frac{\lambda+\ii}{\lambda-\ii}\right)^{\frac{N}{2}\sigma_3}\mathbf{Q}_c^{-1},
\end{equation}
where $\mathbf{Q}_{c}=\frac{1}{\sqrt{2}}\begin{pmatrix}1&-\ii\\-\ii&1
\end{pmatrix}.$
Under the above special case, the matrix function $\mathbf{M}^{[N]}(\lambda; x, y, t)$ satisfies the following Riemann-Hilbert problem.
\begin{rhp}
Let $(x, y, t)\in\mathbb{R}^3$ be arbitrary parameters, and $N\in\mathbb{Z}_{>0}$. Then we can find a $2\times 2$ matrix function $\mathbf{M}^{[N]}(\lambda; x, y, t)$ satisfying the following properties:
\begin{itemize}
\item \textbf{Analyticity}: $\mathbf{M}^{[N]}(\lambda; x, y, t)$ is analytic for $\lambda\in\mathbb{C}\setminus \partial D_0$, where $D_0$ is a big circle involving the point $\lambda=\pm\ii$. It takes the continuous boundary values from the interior and exterior of $\partial D_0$.
\item \textbf{Jump condition}: The jump condition in the boundary of $\partial D_0$ are related by
    \begin{equation}
    \mathbf{M}^{[N]}_{+}(\lambda; x, y, t){=}\mathbf{M}^{[N]}_{-}(\lambda; x, y, t)\mathbf{E}\ee^{\ii\xi\omega\sigma_3}\mathbf{E}^{-1}\mathbf{Q}_{c}\left(\frac{\lambda{+}\ii}{\lambda{-}\ii}\right)^{\frac{1}{2}N\sigma_3}\mathbf{Q}_{c}^{-1}\mathbf{E}\ee^{-\ii\xi\omega\sigma_3}\mathbf{E}^{-1}.
    \end{equation}
\item \textbf{Normalization}: $\mathbf{M}^{[N]}(\lambda; x, y, t)=\mathbb{I}+\mathcal{O}(\lambda^{-1}),$ as $\lambda\to\infty.$
\end{itemize}
\end{rhp}
With the aid of Deift-Zhou nonlinear steepest method, the spatial-temporal pattern for the high order lumps for the large $N$ can be carried out.  For the case $t=0$, the asymptotics for the large order rogue waves was derived by Bilman and Miller very recently \cite{BilmanM-21}. The infinite order rogue waves were given in \cite{BilmanLM-20} by combing the Darboux transformation and Riemann-Hilbert method.  As for the large $t$ and $N$, we would like to explore it in the future work.  In the following, we will study the large $t$ asymptotics for the fixed order $N$.

From the solutions of KP equation \eqref{KP} by the Darboux transformation, we know $u$ satisfies the non-vanishing boundary condition $u=1$. Through a simple symmetry, the non-vanishing background can be eliminated by the transformation $u\to u-1, x\to x-3t, t\to t, y\to y$. Thus in the later analysis, all of the lump solutions are changed into the zero background.

 \section{The lump pattern of KP equation}\label{sec4}
In last section, we have constructed the solutions $q^{[N]}(x,y,t)$ by the Darboux transformation, which is given in a determinant form in Eq.(\ref{sigman}).  In this section, we will utilize the determinant formula to analyze the lump pattern. Actually, for the other form of solutions, we can also analyze the aysmptotics. To describe the structure of lump solution more clearly, we would like to use the method provided in the reference \cite{yang2021rogue}.

From the definition of $\tau_{i,j}^{(n)}$ in Eq.(\ref{tau1}), the elements of $\tau$ can be rewritten as
 \begin{equation}
 \begin{split}\label{tauij}
 &\tau_{i,j}^{(n)}=\left.\frac{1}{i!}(\partial_{\ln p^{*}})^{i}\frac{1}{j!}(\partial_{\ln q})^{j}B^{(n)}\right|_{p^{*}=q=1},\\
 &B^{(n)}=\frac{(-1)^{n}}{1-\frac{(\ee^{\ln p^{*}}-1)(\ee^{\ln q}-1)}{(\ee^{\ln p^{*}}+1)(\ee^{\ln q}+1)}}\exp \left(  \sum\limits_{k=0}^{\infty}A^{+}_{2k+1}(\ln p^{*})^{2k+1}+n(\ln p^{*})+\sum\limits_{l=0}^{\infty}A^{-}_{2l+1}(\ln q)^{2l+1}-n(\ln q) \right),
 \end{split}
 \end{equation}
 where $A^{+}_{2k+1}, A^{-}_{2l+1}$ are given in Eq.(\ref{EQ50}). Then we can rewrite the coefficient of $B^{(n)}$ as
 	\begin{equation}
 	 (-1)^{n}\sum\limits_{\nu=0}^{\infty}\left( \frac{(\ee^{\ln p^{*}}-1)(\ee^{\ln q}-1)}{(\ee^{\ln p ^{*}}+1)(\ee^{\ln q}+1)}\right) ^{\nu}=(-1)^{n} \sum\limits_{\nu=0}^{\infty}\left( \frac{\ln p^{*}\ln q}{4}\right) ^{\nu}\exp\left( \nu \sum\limits_{j=1}^{\infty}s_{j}\left( (\ln p^{*})^{j}+(\ln q)^{j}\right) \right),
 	\end{equation}
 	where $s_{j}$ is given by the following expansion:
 	\begin{equation}\label{EQ65}
 		 \sum_{j=1}^{\infty} s_{j} \lambda^{j}=\ln \left[\frac{2}{\lambda} \tanh \left(\frac{\lambda}{2}\right)\right].
 	\end{equation}
 	It is clear that $s_{2j+1}=0, j=0,1,\cdots$. Next, we introduce some new notations for simplification:\djy{
 	 \begin{equation}\label{EQ70}
 	\begin{split}
 	x_{1}^{+}(n)&=A_{1}^{+}+ n=(x+3 t) + 2 \ii y + n+3t+a_{1},\\
 	x_{1}^{-}(n)&=A_{1}^{-}- n=(x+3 t) - 2 \ii y - n+3t+a^{*}_{1},\\
 	x_{2k+1}^{+}(n)&=A^{+}_{2k+1}=\frac{(x+3 t)+2^{2k+1} \ii y}{(2k+1) !}+\frac{3^{2k+1}t }{ (2k+1)!}+a_{2k+1},\\ x_{2k+1}^{-}(n)&=A^{-}_{2k+1}=\frac{(x+3 t)-2^{2k+1} \ii y}{(2k+1)!}+\frac{3^{2k+1} t}{(2k+1)!}+a^{*}_{2k+1},\quad k\geq 1,\\
 	\end{split}
 	\end{equation}}
 %这里只是将原先的式子化简了，本质上没有改变
where  $a_{2k+1}=\hat{a}_{2k+1}^{*} \left( k=1, 2, \cdots, N-1\right)$ are given in Eq.(\ref{EQ45}). Without loss of generality, we can assume $a_{1}=0$, then $B^{(n)}$ can be rewritten as
 	\begin{equation}
 		(-1)^{n}\sum\limits_{\nu=0}^{\infty}\left( \frac{\ln p^{*}\ln q}{4}\right) ^{\nu}\exp\left( \nu \sum\limits_{j=1}^{\infty}s_{2j}\left( (\ln p^{*})^{2j}+(\ln q)^{2j}\right) +\sum\limits_{k=0}^{\infty}x_{2k+1}^{+}(\ln p^{*})^{2k+1}+\sum\limits_{l=0}^{\infty}x_{2l+1}^{-}(\ln q)^{2l+1}\right).
 	\end{equation}
 	
 	%Considering the Taylor expansion of $B^{(n)}$, we find that  $\tau_{i,j}^{(n)}$ in  Eq.(\ref{tauij}) is  the coefficient of  $(\ln p^{*})^{i}(\ln q)^{j}$ in the Taylor expansion and   we can rewrite it as a Schur polynomial:
 Considering the Taylor expansion of $B^{(n)}$ in Eq.\eqref{tauij}, we find that the coefficient $\tau_{i,j}^{(n)}$ of $(\ln p^{*})^{i}(\ln q)^{j}$ is related to the Schur polynomial:
 	 \begin{equation}\label{EQ62}
 	\tau_{i, j}^{(n)}=(-1)^{n}\sum\limits_{\nu=0}^{\min (i,j)} \frac{1}{4^{\nu}} S_{i-\nu}\left(\mathbf{x}^{+}(n)+\nu \mathbf{s}\right) S_{j-\nu}(\mathbf{x}^{-}(n)+\nu \mathbf{s}),
 	\end{equation}
 	where  $\mathbf{x}^{\pm}(n)=\left(x_{1}^{\pm}(n),0, x_{3}^{\pm}(n),0, \cdots,0,x_{2k+1}^{\pm}(n),0,\cdots \right)$ and $\mathbf{s}=\left( 0, s_{2}, 0, s_{4}, \cdots,0,s_{2k},0,
 	\cdots \right)$. The  definition of Schur polynomial $S_{j}(\mathbf{x})$ with  $\mathbf{x}=\left( x_{1}, x_{2}, \cdots \right)$ is
 	\begin{equation}
 	\sum_{j=0}^{\infty} S_{j}(\mathbf{x}) \lambda^{j}=\exp \left(\sum_{j=1}^{\infty} x_{j} \lambda^{j}\right).
 	\end{equation}
Afterwards, the solution of KP equation can be expressed by the $\sigma_1$ and $\sigma_0$ function, which is shown in theorem \ref{exact solution}.
%Combing with the equation \eqref{qr}, we obtain the following theorem:
 \begin{theorem}\label{exact solution}
 	\djy{The $N$-th  order lump solutions for the KP equation  are given by
 	\begin{equation}\label{kp solution}
 	u_{N}(x,y,t;\mathbf{A})=\left|\frac{\sigma_{1}(x-3t, y, t; a_3,\cdots,a_{2N-1})}{\sigma_{0}(x-3t, y, t; a_3,\cdots,a_{2N-1})}\right|^{2}-1,
 	\end{equation}
	where $N$ is the order of solution and $\sigma_{n}$ is given by Eq.(\ref{sigman})(\ref{EQ65})(\ref{EQ70})(\ref{EQ62}). $\mathbf{A}=(a_3,a_5,\cdots,a_{2N-1})$ and $a_3,\cdots,a_{2N-1}$ are some free parameters in Eq.(\ref{EQ70}).}
 \end{theorem}
 \begin{remark}\label{remark:symmetry}
From the definition $\tau$ function in Eq.\eqref{EQ62}, if all the parameters $a_{2k+1}, (k=1,2,\cdots, 2N-1)$ are purely imaginary number, then we have $\mathbf{x}^{\pm}(n)\Big|_{x\to-x, t\to -t}=-\mathbf{x}^{\mp}(n)$. With a simple calculation, we get a symmetry relation $\tau_{2i-1,2j-1}^{(n)}\Big|_{x\to-x, t\to-t}=\tau_{2j-1,2i-1}^{(n)}$, spontaneously, the solution $u$ will satisfy $u_{N}(x, y, t;\mathbf{A})=u_{N}(-x, y, -t; \mathbf{A})$.
%From the definition $\tau$ function in Eq.\eqref{EQ62}, if all the parameters $a_{2k+1}, (k=1,2,\cdots, 2N-1)$ are pure imaginary number, then we have $x_{1}^{+}(n)\Big|_{x->-x, t->-t}=-x_1^{-}(n), x_{2k+1}^{+}(n)\Big|_{x->-x, t->-t}=-x_{2k+1}^{-}(n)$. With a simple calculation, we get a symmetry relation$\tau_{2i-1,2j-1}^{(n)}\Big|_{x->-x, t->-t}=\tau_{2i-1,2j-1}^{(n)}$, under this choice, the solution $u$ also satisfies $u_{N}(x, y, t, \mathbf{A})=u_{N}(-x, y, -t, \mathbf{A}), $ where $\mathbf{A}=(a_3,a_5,\cdots,a_{2N-1}).$
\end{remark}
	\begin{remark}
		With the method in \cite{zhang2020higher}, we know that the high order lump solution can also be expressed into another equivalent formula \djy{$u_{N}(x, y, t;\mathbf{A})=\frac{\partial^{2}}{\partial {x}^2}\ln(\sigma_{0}(x-3t, y, t;\mathbf{A}))$.}
	\end{remark}

\subsection{ The classification of lump solution and lump pattern}\label{4.1}
%Similar to the high order rogue waves in NLS equation, the lump solution of KP equation is also a rational solution.
{In theorem \ref{exact solution}, we have derived the high order lump solution for KP equation, next we begin to analyze the asymptotics or the spatial-temporal pattern for these high order lump solutions. It can be seen that the solution in theorem \ref{exact solution} contains some free parameters $a_{2k+1}, (1 \leqslant k\leqslant N-1 )$. In recent literature \cite{yang2021rogue}, the authors discussed the asymptotics about these parameters in rogue wave pattern for NLS equation. Unlike to the rogue waves, the lump solution involves three variables $x, y$ and $t$, where the added variable $y$ is similar to the time variable $t$ in $(1+1)$ dimensional system, that is, we can regard the variable $t$ as a new one. Then the asymptotics to lump solutions will be diverse, we can not only  study the asymptotics with respect to $t$ but also about the internal parameters $a_{2k+1}$. By a tedious calculation, we give the main propositions and the theorems about the asymptotic analysis, and their proof is put in the appendix.}

\begin{proposition} \label{property1}
	%这个地方在when t is large后面加了一点and satisfy是为了证明需要
\djy{	When $t$ is large and $|t|\gg |a_{2k+1}|^{\frac{3}{2k}}, (1\leqslant k \leqslant N-1)$, if $(x,y)$ satisfies $\sqrt{(x+3 t)^{2}+4y^{2}}=\mathcal{O}(|t|^{\frac{1}{3}})$,} then the $N$-th lump solution \djy{$u_{N}(x,y,t;\mathbf{A})$} in Eq.(\ref{KP}) decays to the zero background, except at or near the point $\left(x, y, t\right)=\left(x_{0}, y_{0},t\right)$, where
	\begin{equation}\label{root}
	x_{0}+3 t+2\ii y_{0}=A^{\frac{1}{3}} z_{0}, \quad A=-3 t,
	\end{equation}
	\jy{
	\noindent{ and $z_{0}$ is the non-zero root of Yablonskii-Vorob'ev polynomial $Q_{N}^{[1]}(z)$}.}
\end{proposition}
The Yablonskii-Vorob\'ev polynomial, which is originated from the rational solutions of the second Painlev\'e equation ($P_{II}$) \cite{vorob1965rational,yablonskii1959rational},
\begin{equation}
w''=2w^3+zw+\alpha.
\end{equation}
When $\alpha$ is an integer, this equation has the rational solutions as \djy{
\begin{equation}
w(z; N)=\frac{d}{dz}\ln\frac{Q_{N-1}^{[m]}(z)}{Q_{N}^{[m]}(z)},\quad N\geq 1,\quad  w(z, 0)=0,\quad w(z; -N)=-w(z; N),
\end{equation}
where $m$ is a non-negative integer and $Q_{N}^{[m]}(z)$ is called the Yablonskii-Vorob'ev polynomial.} Moreover, it is equivalent to a determinant composed by the Schur polynomial $p_{k}^{[m]}(z)$ \cite{kajiwara1996determinant}:
\begin{equation}\label{eq p}
\begin{split}
{Q}_{N}^{[m]}(z)&=c_{N}\underset{1\leqslant i,j\leqslant N}{\det}\left[p_{2i-j}^{[m]}(z)\right],\quad c_{N}=\prod_{j=1}^{N}(2 j-1) ! !,\\
\sum\limits_{k=0}^{\infty} p_{k}^{[m]}(z) \lambda^{k}&=\exp \left(z \lambda-\frac{2^{2m}}{2m+1} \lambda^{2m+1}\right),\quad
p_{k}^{[m]}(z)=0,\,\,\, (k<0).\\
\end{split}
\end{equation}
In the later asymptotic analysis, we find that the locations of lump solution have an intimate relationship with the root structures of the Yablonskii-Vorob'ev polynomial $Q_{N}^{[m]}(z)$, which has been studied in \cite{balogh2016hankel,clarkson2003second, fukutani2000special, taneda2000remarks,buckingham2014large}.  With the result in \cite{fukutani2000special}, we know the order of $Q_{N}^{[m]}(z)$ is $N(N+1)/2$, among which the number of \djy{nonzero roots} is $N_{p}^{[m]}$ and the multiplicity of zero root is $N_{0}^{[m]}$, where
\begin{equation}\label{N02}
{N}_{0}^{[m]}=\left\{\begin{array}{cl}
N \bmod (2m+1),\quad &0\leqslant N \bmod (2m+1)\leqslant m , \\[3pt]
2m-\left( N \bmod \left( 2m+1 \right) \right),\quad &N \bmod (2m+1) > m,
\end{array}\right.
\end{equation}
and
\begin{equation}\label{Npm}
N_{p}^{[m]}=\frac{1}{2}\left[N(N+1)-N_{0}^{[m]}\left({N}_{0}^{[m]}+1\right)\right].
\end{equation}
In addition, in \cite{clarkson2003second}\cite{fukutani2000special}, the authors have proved that the \djy{nonzero roots} of the Yablonskii-Vorob'ev polynomial are all simple.

Furthermore, in the neighbourhood of the special point $ (x,y)=\left(  x_{0}, y_{0}\right) $, the asymptotics will be different, which is shown in proposition \ref{property2}.
\begin{proposition}\label{property2}
	\djy{ When $t$ is large and $|t|\gg |a_{2k+1}|^{\frac{3}{2k}}, (1\leqslant k \leqslant N-1)$, in the neighbourhood of the point $(x,y)=(x_{0},y_{0})$ with $\sqrt{\left(x-x_{0}\right)^{2}+4 \left(y-y_{0}\right)^{2}}=\mathcal{O}(1)$,} where $(x_{0}, y_{0})$ is given in proposition \ref{property1}, then the $N$-th order lump solution $u_{N}(x,y,t;\mathbf{A})$ in Eq.(\ref{kp solution}) approaches \djy{a  first-order lump} $u_{1}\left( x-\left( x_{0}+3t\right) ,y-y_{0},t\right) $, where
	\begin{equation}
	u_{1}(x, y, t)=\frac{-32(x+3t)^2+128y^{2}+8}{\left( 4 \left( x+3t\right) ^{2} +16 y^{2}+1\right) ^{2}}+\mathcal{O} \left(\left|t \right|^{-\frac{2}{3}}\right).
	\end{equation}
	 From this expression, it is easy to see that the velocity about $u_{1}$ is $v=(v_x,v_y)$, where $v_x=-3t, v_y=0$.
\end{proposition}
What is more, the asymptotic behavior in the near point $(x,y)=(-3t,0)$ must be different, whose asymptotics is given in proposition \ref{property3}.
\begin{proposition}\label{property3}
   \djy{	When $t$ is large and $|t|\gg |a_{2k+1}|^{\frac{3}{2k}}, (1\leqslant k \leqslant N-1)$,} in the neighborhood of $(x,y)=(-3t,0)$ with
	$\sqrt{(x+3 t)^{2}+4y^{2}}=\mathcal{O}(1)$, the  $N$-th order lump solution $u_{N}(x,y,t;\mathbf{A})$ in Eq.(\ref{kp solution}) will asymptotically approach a lower $N_{0}^{[1]}$-th order lump $u_{N_{0}^{[1]}}(x,y,t)$. In this case, $N_{0}^{[1]}$ is either 1 or 0 with the definition in Eq.(\ref{N02}).  Thus we have
	\begin{equation}\label{eq50}
	u_{N_0^{[1]}}(x, y, t)=\left\{\begin{array}{cl}
	0, &N_{0}^{[1]}=0, \\[8pt]
	\frac{-32\left( x+3t\right) ^2+128y^{2}+8}{\left( 4 \left( x+3t\right) ^{2} +16 y^{2}+1\right) ^{2}},& N_{0}^{[1]}=1.
	\end{array}\right.
	\end{equation}
\end{proposition}
With the result in proposition \ref{property2}, \ref{property3}, \djy{when $t$ is large, the whole asymptotics contains two different types--in the neighborhood of $(x,y)=(x_ {0}, y_ {0})$ or $(x,y)=(-3t,0)$.} In total, the asymptotic expression about the KP equation is given in theorem \ref{theo2}.
\begin{theorem}\label{theo2}
	According to  proposition \ref{property1}, \ref{property2}, \ref{property3}, we can get a conclusion that when $t$ is large and satisfies $|t|\gg |a_{2k+1}|^{\frac{3}{2k}}, (1\leqslant k \leqslant N-1)$,
%under the condition $a_{2k+1}$,($1\leqslant k\leqslant N-1)=\mathcal{O}(1)$,
the asymptotic expression about the lump soliton of KP equation is
	\begin{equation}\label{uN}
	u_{N}(x, y, t;\mathbf{A})\to u_{N_0^{[1]}}(x, y, t)+\sum_{j=1}^{N_{p}^{[1]}}\left(\frac{-32\left( x-x_{0}^{(j)}\right) ^{2}+128\left( y-y_{0}^{(j)}\right) ^{2}+8}{\left(4\left(x-x_{0}^{(j)} \right)^{2}+16\left(y-y_{0}^{(j)}\right)^{2}+1\right)^{2}}\right),
	\end{equation}
	where $N_{0}^{[1]}, N_{p}^{[1]}, (x_{0},y_{0})$ and $u_{N_{0}^{[1]}}(x,y,t)$ are given in Eq.(\ref{N02}), Eq(\ref{Npm}), Eq.(\ref{root}) and Eq.(\ref{eq50}) respectively. The superscript $^{(j)}$ means the $j$-th $(x_{0},y_{0})$.
\end{theorem}
\jy{In fact, the number of the \djy{ first-order lumps} is always $N(N+1)/2$ when  $t$ is large no matter $N_{0}^{[1]}$ is 0 or 1}.

%Under the condition of all internal parameters $a_{2k+1}$ are $\mathcal{O}(1)$, we give the asymptotic expression about the lump solution as $t$ is large.
In the following proposition \ref{property4}, we will analyze the asymptotics with respect to the internal parameter $a_{2k+1}$. \djy{By the calculation, we find an approximate critical point $\frac{2^{2k}}{2k+1}$ about the parameters  $a_{2k+1}$, and the behaviors of lump solution will change essentially whether $a_{2k+1}$ is greater than $\frac{2^{2k}}{2k+1}$ or not. }
\begin{proposition}\label{property4}
Suppose only one parameter $a_{2m+1}$ large enough with \djy{$|a_{2m+1}|\gg \frac{2^{2m}}{2m+1}, (1<m\leq N-1)$ and other parameters satisfying $a_{2k+1}=o(|a_{2m+1}|^{\frac{2k}{2m+1}}),(k\neq m)$,} then the asymptotics can be summarized in the following aspects.
\begin{itemize}
\item   When $|t| \gg |a_{2m+1}|^{\frac{3}{2m}}$, the $N$-th lump solution is still decomposed into  $N(N+1)/2$ \djy{first-order lumps}, whose asymptotic expression can also be expressed by Eq.(\ref{uN}).
	
\item  When $|t| \ll |a_{2m+1}|^\frac{2}{2m+1}$, as $(x,y)$ is far away from the point $(-3t,0)$, the $N$-th order lump solution
%这里两个内部参数故意没有换成A
 \djy{$u_{N}(x,y,t;a_3,a_5,\cdots,a_{2N-1})$} will be separated into $N_{p}^{[m]}$ \djy{first-order lumps},  where $N_{p}^{[m]}$ is given in Eq.(\ref{Npm}). That is,   in the neighborhood of $(x, y)=(x_{0}^{[m]},y_{0}^{[m]})$ with $\sqrt{(x-x_{0}^{[m]})^2+4(y-y_{0}^{[m]})^2}=\mathcal{O}(1)$, the  lump solution  asympotically approaches a \djy{first-order lump}
 ${u}_{1}\left(x-(x_{0}^{[m]}+3t), y-y_{0}^{[m]},t\right)$, where
		\begin{equation}\label{transfomation2}
		x_{0}^{[m]}={\rm Re}\left( \left( -\frac{2m+1}{2^{2m}}a_{2m+1}\right) ^{\frac{1}{2m+1}}z_{0}^{[m]}\right)-3t,\quad y_{0}^{[m]}=\frac{1}{2}{\rm Im}\left( \left( -\frac{2m+1}{2^{2m}}a_{2m+1}\right) ^{\frac{1}{2m+1}}z_{0}^{[m]}\right),
		\end{equation}
and $z_{0}^{[m]}$ is the nonzero root of $Q_{N}^{[m]}(z)$; additionally, in the neighborhood of the point $(x,y)=(-3t,0)$ with $\sqrt{(x+3t)^2+4y^2}$ $= \mathcal{O}(1)$, the $N$-th order lump solution  will asymptotically approaches a lower $N_{0}^{[m]}$-th order lump solution \djy{$u_{N_{0}^{[m]}}(x,y,t;a_3,a_5,\cdots,a_{2N_{0}^{[m]}-1})$}, where $N_{0}^{[m]}$ is given in Eq.(\ref{N02}).
\end{itemize}
\end{proposition}

\subsection{The classification of high order lumps with purely imaginary parameters $\mathbf{A}$}
In the last subsection, we give a detailed asymptotic analysis for the  lump solutions of KP equation in proposition \ref{property1}-\ref{property4}, which can be utilized to give the whole scenery of lumps. In this subsection, we would like to use these properties to give the classification of these lump solutions.
	% In this subsection, we would like to  give the classification of  these lump solutions and analyze the whole scenery of different types of lump solutions utilizing these propositions.
Compared to the rogue wave in NLS equation, the lump solution of KP equation is $1+2$ dimensional and has a two-dimensional dynamic graph as the variation of time variable $t$. If $t$ is large enough, through the proposition \ref{property1}-\ref{property3}, we know that the patterns of lumps have the similar structure. But when $t$ is small, the lump patterns will be diverse. Based on the result in remark \ref{remark:symmetry}, we can classify the lump solution if we give a constraint to these parameters $a_{2k+1}, (k=1,2,\cdots, N-1)$ and set them all purely imaginary. Under this choice, the evolution process is symmetry on $t=0$ when $x$ is changed into $-x$, then we can deem that the lumps have the strongest interaction at $t=0$, which is the reason of classification.  At this time, the behavior of lump solution is consistent with the rogue wave in \cite{yang2021rogue,Ling2013PRE}, thus we would like to classify the lumps into the following four categories by the dynamics of solution at $t=0$:
\begin{itemize}
	\item {\bf Complete polymerization type:} $u_N(x,y,0;\mathbf{A})$ has the maximal peak at $(x,y)=(0,0)$.
	\item {\bf Partially polymerization type:} the asymptotic state of $u_N(x,y,0;\mathbf{A})$  has a lower order lumps with the maximal peak $u_{K}(x,y,0;\mathbf{A}=\mathbf{0})$, $2\leq K\leq N$.
	\item {\bf Completely separating type:} the asymptotic state of $u_N(x,y,0;\mathbf{A})$ is separately distributed with the first order lumps $u_1(x-x_i,y-y_i,0)$, $i=1,2,\cdots, N(N+1)/2$, where $(x_i,y_i)$s are the central points of the first lumps. The least distance $d_C$ between any two central points $(x_i,y_i)$ of these first lumps is enough wide (in this work, we set the criterion $d_C\geq 1$).
	\item {\bf Hybrid type:} the other case are not involved in the above three categories.
\end{itemize}

By the above analysis and the determinant formula, we know that the spatial distribution of lumps at $t=0$ is affected by the parameters $a_{2k+1}, (k=1,2,\cdots, N-1)$. In general, the properties for $u_N(x,y,0;\mathbf{A})$ with $\mathbf{A}$ purely imaginary are not completely determined. But for one large internal parameter and multiple internal parameters, the distribution can be almost determined by the roots of $Q_{N}^{[m]}(z)$ as given in the reference \cite{yang2021rogue}.  Now, we give a description about the classification.

{\bf Case 1: Completely polymerization type}

By the proposition \ref{prop1} and Darboux matrices \eqref{DT-decom}, we can prove that the lump solutions in Eq.(\ref{kp solution}) will attain the maximal value $u_{N}(0,0,0;\mathbf{0}) = (2N + 1)^2-1$ at the point $(x,y,t)=(0,0,0)$ by choosing $\mathbf{A}=\mathbf{0}$, which will yield the high order lumps with completely polymerization type. One of calculation way for $u_{N}(0,0,0;\mathbf{0})$ was given in \cite{WangYWH17}. On the other hand, as $t\to\pm\infty$, by the proposition \ref{property2} and \ref{property3}, we know that high order lumps can be decomposed into $\frac{1}{2}N(N+1)$ first-order lumps, whose locations are determined by the roots of $Q_{N}^{[1]}(z)$. We find that $u_{N}(0,0,0;\mathbf{0}) = (2N + 1)^2-1=8\times \frac{1}{2}N(N+1)$, which means that all the first order lumps will collide at the origin, where $8$ is the height of first order lump.

\begin{figure}[!h]
		\centering
		\includegraphics[width=\linewidth]{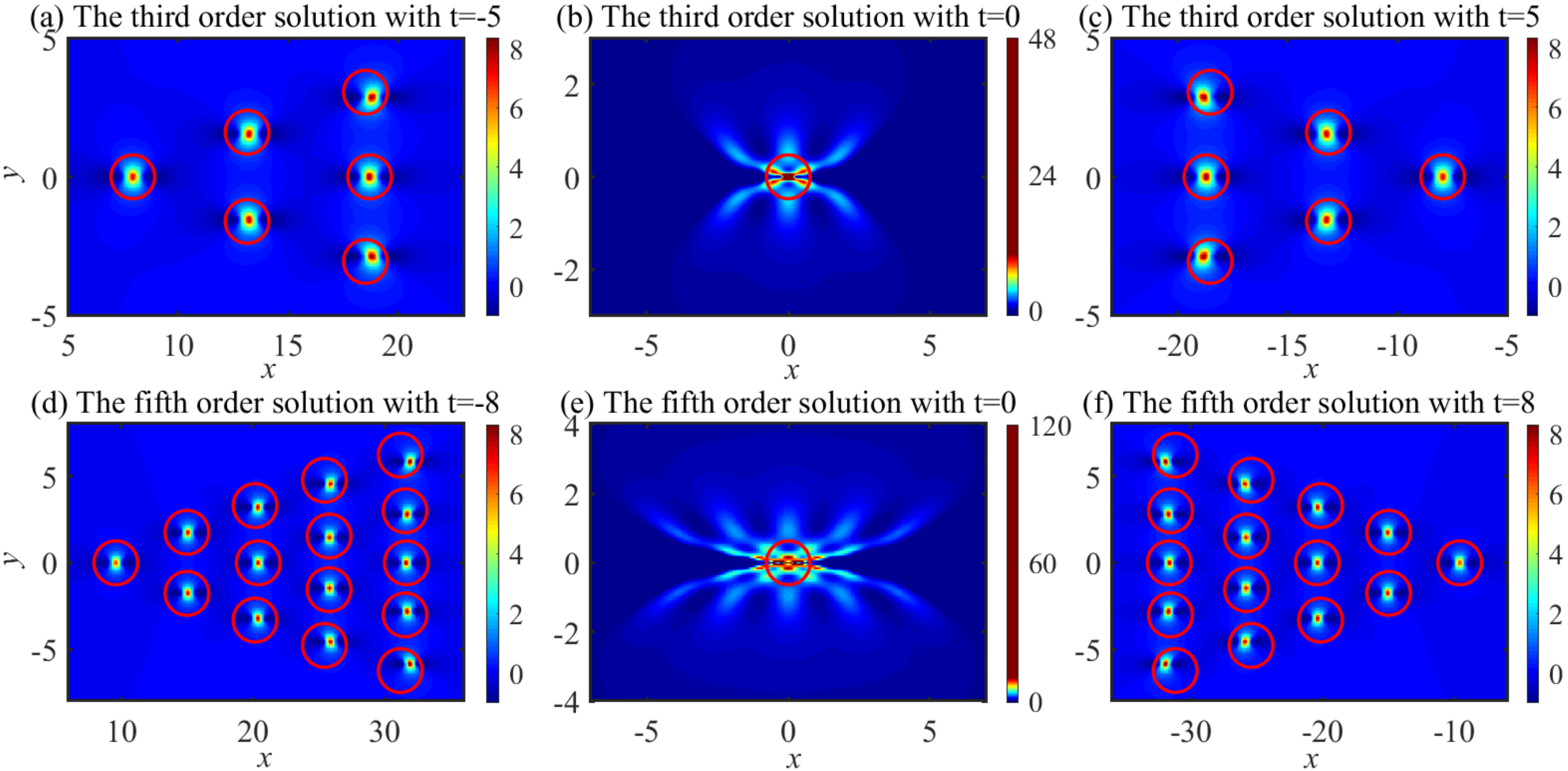}
		\caption{The evolution lump solutions of KP equation  when $N=3,5$ and $\mathbf{A}=\mathbf{0}$. The red circles are the predicated locations by the roots of $Q_{3}^{[1]}$ and $Q_{5}^{[1]}$ respectively.}
%Ture solutions are plotted with Theorem \ref{theo2} and the  red circles are predicted  locations of \djy{ these first-order lumps using} Eq.(\ref{transformation}). The pictures in the first row are the third order lump solution $u_{3}(x,y,t;0,0)$  and the pictures in the second  row are the fifth order lump solution $u_{5}(x,y,t;0,0,0,0)$.  When $t=0$, the maximum of (b) can reach \djy{ $48$ and (e) can reach $120$. }}
		\label{true and predict}
	\end{figure}

Especially, we would like to exhibit the dynamic behavior by the third and fifth-order lumps (Fig. \ref{true and predict}). Firstly, we analyze the dynamics of high order lumps with large $t$. For $N=3$, by the formula \eqref{eq p}, we have
	\begin{equation}\label{Q3}
	Q_{3}^{[1]}(z)=z^{6}+20z^{3}-80,
	\end{equation}
which implies that the corresponding locations of lumps $(x_0,y_0)$ are given by Eq.(\ref{root}):
	\begin{equation}\label{transformation}
	x_{0}={\rm Re}((-3t)^{\frac{1}{3}}z_{0})-3t,\quad y_{0}={\rm Im}((-3t)^{\frac{1}{3}}z_{0})/2.
	\end{equation}
The case for $N=5$ can be calculated similarly.
Obviously, the locations of lumps are related to the time variable $t$.
It can be seen that when $t$ is large enough, the third and fifth order lump solution will be separated into $6$ and $15$ \djy{first-order lumps} respectively. When $t=0$, all the first-order lumps collide at the origin $(x,y,t)=(0,0,0)$ and form a maximal peak. We can find that the peak's values of the third and fifth order lumps are $6\times 8=48$ and $15\times 8=120$ respectively, which exhibits the linear superposition for the first order lumps. Under this special setting $\mathbf{A}=0$, the rule holds for all high order lumps.

{\bf Case2: Partially polymerization type}

\begin{figure}[!h]
	\centering
	\includegraphics[width=\linewidth]{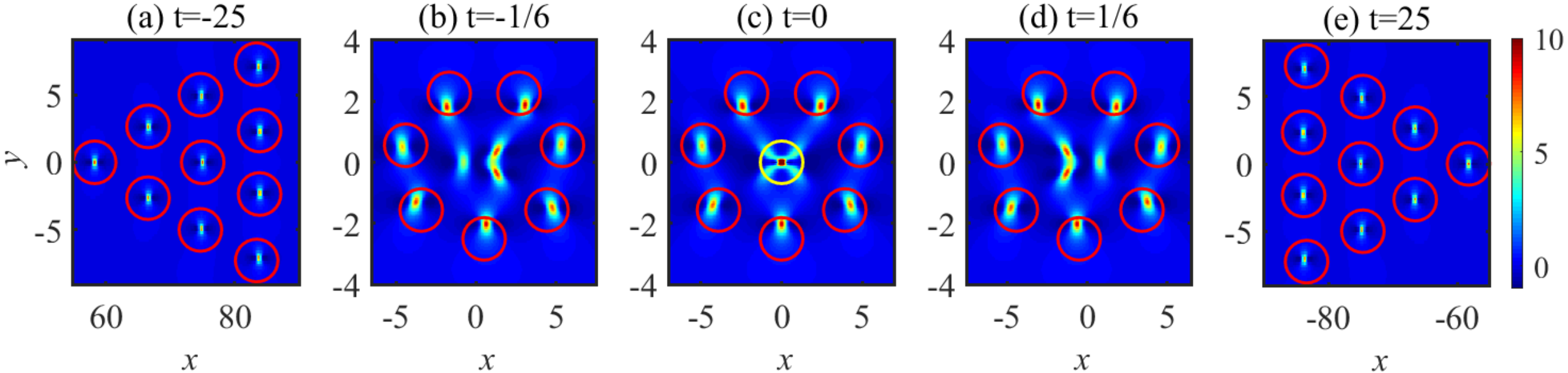}
	\caption{The fourth order lump solution of KP equation by choosing $a_3=a_5=0, a_{7}=50\ii$. \djy{The center of each picture is $(x,y)=(-3t,0)$. These red circles are the predicted  locations of these first-order and the yellow circle gives the predicted location of the second-order  lump. }}
	\label{fig.large a7}
\end{figure}

The partially polymerization lump solutions are defined by the asymptotic state of $u_N(x,y,0;\mathbf{A})$ having a lower order lumps with maximal peak. Under this case, we give a sufficient condition with the parameters choice of $\mathbf{A}$ with one parameter $a_{2m+1}$ large enough. With the aid of proposition \ref{property4}, there will appear a lower order (less than $N$) lump solution at $(x,y,t)=(0,0,0)$. Specifically, we set $\left|a_{2m+1}\right|\gg\frac{2^{2m}}{2m+1}$ and \djy{
\begin{equation*}
\left\{
\begin{split}
a_{2k+1}&=0, \qquad  1\leq k< N_{0}^{[m]},\\
|a_{2k+1}|&\ll|a_{2m+1}|^{\frac{2k}{2m+1}},\qquad  k \geq N_{0}^{[m]},
\end{split}
\right.
\end{equation*}
}where $N_0^{[m]}$ is defined by Eq.\eqref{N02} and $N_{0}^{[m]}\geq 2$. Then the lower order lump solution located at $(x,y)=(0,0)$ can reach the corresponding maximum with its own order.
%the lower order lump solution can reach a maximum, which is matching its own order. Especially, the order is controlled by the multiplicity of zero root of the polynomial $Q_{N}^{[m]}(z)$ defined by Eq.\eqref{N02}. This case, we set the multiplicity greater than one.
%And the properties of these first-order lump solution is similar to the large $t$ situation, whose location can also be calculated by the non-zero root of polynomial $Q_{N}^{[m]}(z)$ defined by Eq.\eqref{eq p}.
By choosing some proper parameters, we exhibit the evolution dynamics in Fig.\ref{fig.large a7}.

In Fig.\ref{fig.large a7}, we set $N=4$ and $a_7=50\ii$. With a simple calculation, we know $m=3, N_{0}^{[3]}=2$ and $N_{p}^{[3]}=7$. At $t=0$, there appear seven first-order lumps embraced with one second-order lump $u_{2}(x,y,0;0)$ located at $(x,y)=(0,0)$. The locations of these seven first-order lumps can be obtained by the nonzero roots of $Q_{4}^{[3]}(z)$ in Eq.(\ref{eq p}) and the transformation Eq.(\ref{transfomation2}). When $t=\pm 1/6$, an asymmetrical second-order lump $u_{2}(x,y,\pm 1/6;0)$ is approximately in the neighborhood of $(x,y)=(\mp 1/2,0)$.

{\bf Case3: Completely separating type}

\begin{figure}[!h]
		\centering
		\includegraphics[width=\linewidth]{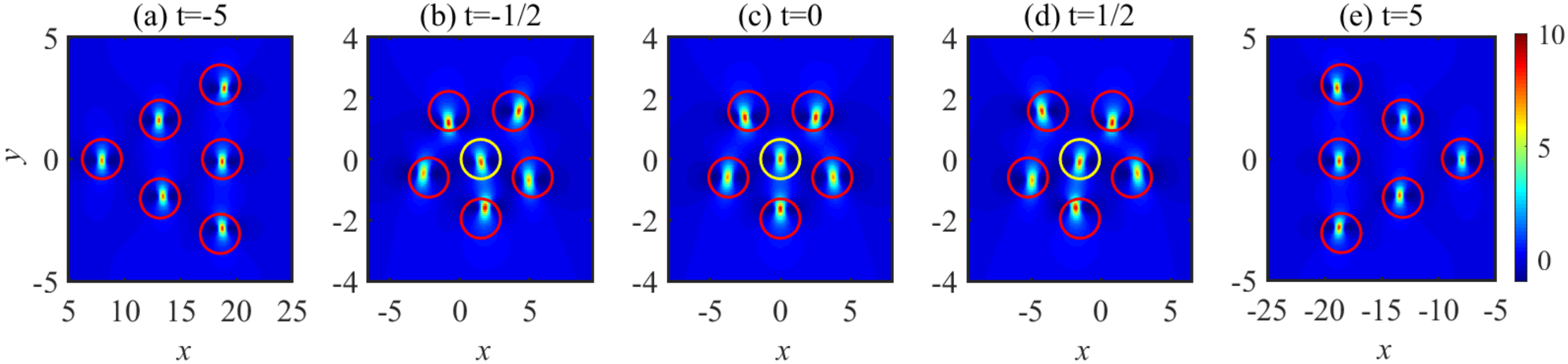}
		\caption{The third order lump solution of KP equation  by choosing $a_{3}=0, a_{5}=20\ii$. \djy{The center of yellow circle  is $(x,y)=(-3t,0)$ and these red circles are the predicted locations by} Eq.(\ref{transfomation2}) and Eq.(\ref{transformation}).}
		\label{fig.large a5}
	\end{figure}

The completely separating lump solutions are defined by the asymptotic state of $u_N(x,y,0;\mathbf{A})$ having the separated first-order lumps. Compared to the sufficient condition given in case 2, we can also give a similar sufficient condition to the parameter $\mathbf{A}$ with one parameter $a_{2m+1}$ large enough, but in this case, $N_{0}^{[m]}$ should be $1$ or $0$. Then there will only appear the first-order lumps at anytime. By choosing some proper parameters, we show one example of this type in Fig. \ref{fig.large a5}.

In Fig.\ref{fig.large a5}, choosing $N=3, a_3=0, a_5=20\ii$, then we get the third-order lump with $m=2, N_{p}^{[2]}=5, N_0^{[2]}=1$. When $t=0$, there appear six first-order lump solutions, and the locations for the five lumps can be obtained by the roots of $Q_{3}^{[2]}(z)$ in Eq.(\ref{eq p}) and the transformation Eq.(\ref{transfomation2}). The left first-order lump is still located in the neighborhood of $(x, y, t)=(0, 0, 0)$.

{\bf Case4: Hybrid type}

Apart from above three cases, we call the other lump solutions as the hybrid type ones. In this case, there exist a lot of choices to the parameters $a_{2k+1}$. And we list some ones as follows:

1. $|a_{2k+1}|<\frac{2^{2k}}{2k+1}, (1\leq k \leq N-1),$ but $\mathbf{A}\neq \mathbf{0}$;

2. $|a_{2m+1}|\gg \frac{2^{2m}}{2m+1}$, but there exists $k$ such that $a_{2k+1}\neq 0$ with $1\leq k< N_{0}^{[m]}$;

3. there are more than two parameters $a_{2m+1}$ large enough.

\begin{figure}[htbp!]
		\centering
		\includegraphics[width=\linewidth]{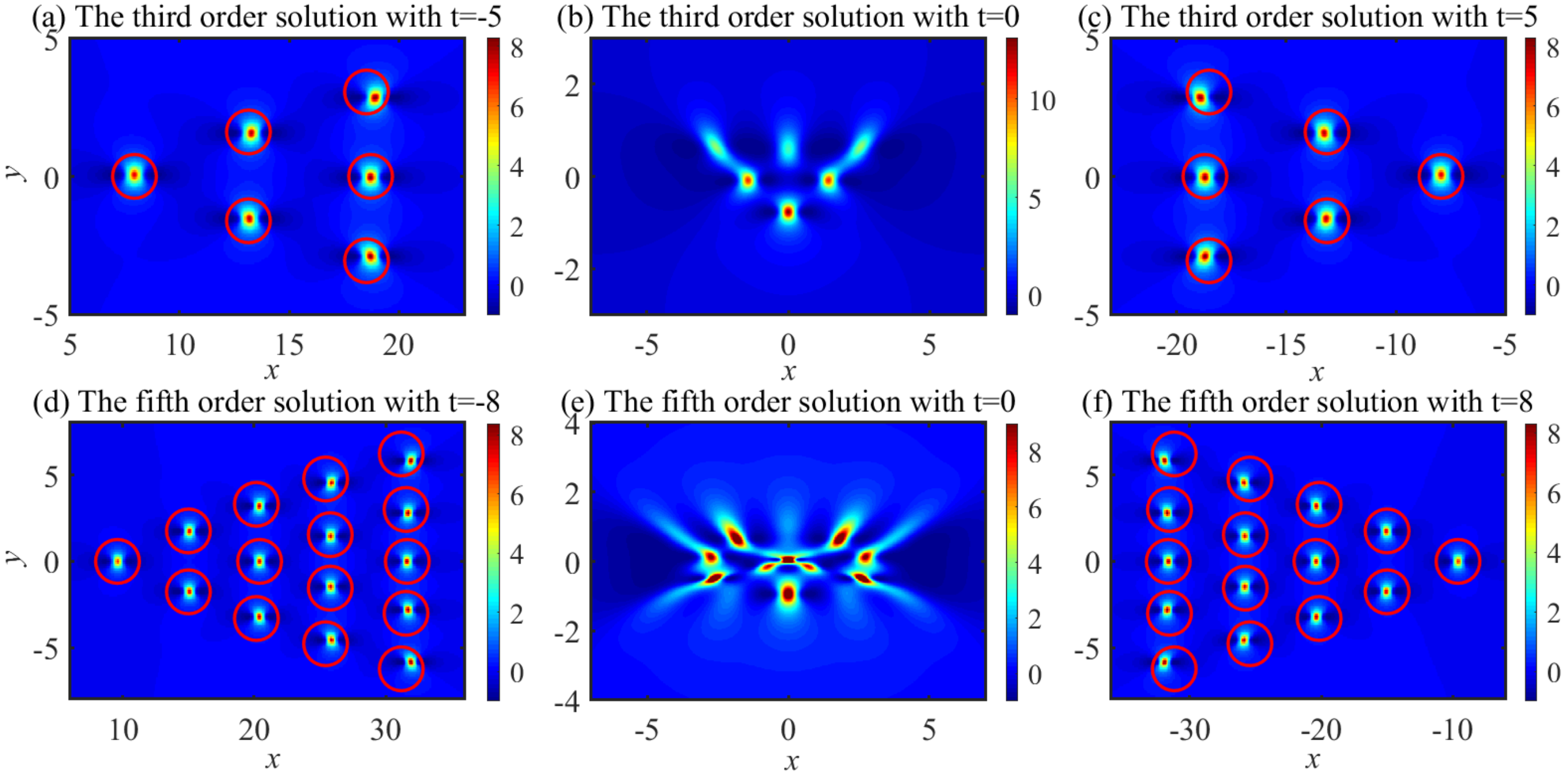}
		\caption{The lump solutions of KP equation  when $N=3,5$, in this case, these parameters $a_{2k+1}<\frac{2^{2k}}{2k+1}$ but $\mathbf{A}  \neq \mathbf{0}$. The red circles are the predicted locations by Eq.(\ref{transformation}). The pictures in the first row are the third order lump solution $u_{3}(x,y,t;\ii,\ii)$  and the pictures in the second  row are the fifth order lump solution $u_{5}(x,y,t;\frac{\ii}{10},\frac{\ii}{5},\frac{\ii}{2},\ii)$. }
		\label{fig.aO1}
	\end{figure}

If these parameters satisfy the first one, then there only appear the first-order lump solutions, but the least distance between two adjacent lumps is too close to distinguish; if these parameters satisfy the second one, there maybe appear a higher order lump solution and some first-order lumps, still, the least distance is too close; if the parameters satisfy the third one, there maybe appear more than two higher order lump solutions, which are similar to the rogue wave in \cite{Ling2013PRE}. We only show the evolutional process in this case by choosing the parameters as the first one, which are shown in Fig.\ref{fig.aO1}.
%From the result of proposition \ref{property1}-\ref{property4}, as $t$ is large, the asymptotics in this case is the same as case 1--case 3, all of them will be separated into $N(N+1)/2$ first-order lump solution.

From Fig. \ref{true and predict} to Fig. \ref{fig.aO1}, we show four types of lump solutions by choosing different parameters. When $t=0$, different types of lumps have different behaviors. But when $t$ is large, all of them have a similar geometric distribution, which can be seen from Fig.\ref{true and predict}(a), Fig.\ref{fig.large a5}(a) and Fig.\ref{fig.aO1}(a). This incredible phenomenon is very strange and we can not give a reasonable explanation about it so far, we just can say that it is determined by the choice of these parameters. In other words, we can regard these parameters as a small perturbation when $t$ is large, this strange phenomenon reflects a fact that if a solution has the same initial data but with a small perturbation, their evolution process would be completely different.

%diverse lump patterns by choosing different parameters and analyze two types of asymptotics with respect to the variable $t$ and the parameter $a_{2m+1}$, which presents richer character than the rogue wave in $(1+1)$ dimensional NLS equation.

\begin{remark}
In the references \cite{Matveev1991DT,ablowitz2000novel,Chang2018KP}, the authors give the lump solution of KP equation via the binary Darboux transformation and the $\tau$ function method, and they give the asymptotics for the lump solutions as $t$ is large. Our method is different from theirs and has its merit. In \cite{ablowitz2000novel,Chang2018KP}, as $t$ is large, the locations of the peak is related to the real root of an orthogonal polynomials, while in our results, the locations are determined by the root of Yablonskii-Vorob'ev polynomial. More importantly, the maximal peak of our higher order lump can be obtained by choosing proper parameters. Apart from the asymptotics with $t$, we also analyze the asymptotics with respect to the parameter $a_{2k+1}$.
\end{remark}

\section{Discussions and conclusions}
In this paper, we give a detailed analysis for the lump pattern on both theoretically and numerically via the one-constraint method, to our best knowledge, there are seldom articles about the asymptotics to the high dimensional system. The lump solution in the KP equation is similar to the rogue wave in the $(1+1)$ dimensional system, which is a hot topic in recent years. Thus the dynamical analysis to the lump solution is extremely significant. With the one-constraint method, we classify the lump patterns into four different types according to the behavior at $t=0$. Observing the $\tau$ function in Eq.\eqref{EQ62}, we find that this solution has a good symmetry if all parameters $a_{2k+1},(k=1,2,\cdots, N-1)$ are purely imaginary. Based on this fact, we can give this kind of classification, while if these parameters are arbitrary complex number, the symmetry of this solution is ambiguous, we have not found a location that can be used as the criterion for classification, this interesting topic can be studied in future.

In this paper, we only use the one-constraint condition. Apart from this, the equations derived from the two-constraint or the three-constraint are also diverse, such as in \cite{cheng1991constraint}, the authors gave a $3\times 3$ matrix eigenvalue Yajima-Oikawa equation from the two-constraint. Moreover, for a general $k$-constraint, there are many high dimensional equations, which can also be reduced to a simple (1+1) dimensional equation. Similar to the one-constraint equation, our idea can be extended to the high order Lax pair and analyze the properties effectively. In \cite{zhang2020higher,ZhangLY-21}, with the Darboux transformation, the authors analyzed the asymptotics for the multi-component system, whose calculation is more complicated. Our method can also be applied to this similar multi-component system and give the corresponding solutions with the Darboux transformation and the $\tau$ fucntion, which can be used to analyze the asymptotics efficiently, if possible, the location of the multi-component system can be determined by the root of a polynomial.

With the theory of inverse scattering, from the Darboux matrix of high dimensional AKNS system, we construct the corresponding Riemann-Hilbert problem. More importantly, this Riemann-Hilbert problem can be used to study the asymptotics when $N$ is large. In the future study, we prepare to analyze the asymptotics of lump solution for KP equation when $N$ is large by using this Riemann-Hilbert problem.
%Additionally, in this paper, we only give a special solutions when $\lambda_1=-\ii$, if $\lambda_1$ is an arbitrary complex number, what is the dynamics of solutions for KP equation will be, we will study the corresponding problem in the next work.
\section*{Acknowledgement}
Liming Ling is supported by the National Natural Science Foundation of China (Grant Nos. 11771151, 12122105), the Guangzhou Science and Technology Program of China (Grant No. 201904010362); Xiaoen Zhang is supported by the National Natural Science Foundation of China (Grant No. 12101246), the China
Postdoctoral Science Foundation (Grant No. 2020M682692).
\section*{Appendix}
In the appendix, we give the proofs about the above propositions and theorems in subsection \ref{4.1}.  Actually, the following proofs follow the calculations in the Ref. \cite{yang2021rogue} with a minor revision on the variables.

\textbf{\emph{Proof of Proposition \ref{property1}.}}
Firstly, we set $\epsilon=|t|^{-\frac{1}{3}}$.  From the definition of $\sigma_{n}$, we can give  the element $\tau_{2i-1,2j-1}^{(n)} $ of $\sigma_{n}(x-3t,y,t;\mathbf{A})$ as
\begin{equation}
	\tau_{2i-1,2j-1}^{(n)}=(-1)^{n}\sum\limits_{\nu=0}^{\min (2i-1,2j-1)} \frac{1}{4^{\nu}} S_{2i-1-\nu}\left(\mathbf{x}^{+}(n)+\nu \mathbf{s}-\mathbf{t}\right) S_{2j-1-\nu}(\mathbf{x}^{-}(n)+\nu \mathbf{s}-\mathbf{t}),
\end{equation}
where $\mathbf{t}=\left( 3t,0,\frac{3t}{3!},0,\cdots,0,\frac{3t}{(2k+1)!},0,\cdots\right) $. \djy{Under the condition $ |a_{2k+1}| \ll |t|^{\frac{2k}{3}}$, which indicates $a_{2k+1} \epsilon ^{2k+1} =o(\epsilon)$,}  when $\sqrt{(x+3 t)^{2}+4y^{2}}=\mathcal{O} \left( |t|^{\frac{1}{3}} \right)=\mathcal{O}\left(\epsilon^{-1}\right)$,
	we have
	\begin{equation}
	\begin{split}
	S_{k}\left(\textbf{x}^{+}(n)+ \nu \textbf{s}  -\mathbf{t}\right) &=S_{k}\left(x+3 t+2 \ii y+n, \nu s_{2} , \ldots\right) \\
	&=\epsilon^{-k} S_{k}\left((x+3 t+2 \ii y+n) \epsilon, \nu s_{2} \epsilon^{2},\left(\frac{x+3 t+2^{3} \ii y}{3 !}+a_{3}+\frac{(3^{3}-3) t}{3 !}\right) \epsilon^{3}, \cdots\right) \\
	&=\epsilon^{-k} S_{k}\left((x+3 t+2 \ii y) \epsilon, 0, \frac{\left(3^{3}-3\right) t}{3 !} \epsilon^{3}, 0, 0,0,\cdots\right)[1+\mathcal{O}(\epsilon)] \\
	& \sim S_{k}(x+3 t+2 \ii y, 0,4 t, 0,0,0,\cdots).\\
	\end{split}
	\end{equation}
	Thus  the Schur polynomial can be given as
	\begin{equation}
	S_{k}\left(\textbf{x}^{+}(n)+ \nu \textbf{s} -\mathbf{t} \right) \sim S_{k}\left(\mathbf{v}\right),\quad |t|\to \infty
	\end{equation}
	where
	\begin{equation}
	\begin{split}
	&\mathbf{v}=\left( x+3 t+2 \ii y, 0,4 t, 0,0,0, \cdots\right),\\
	&\sum\limits_{k=0}^{\infty} S_{k}\left(\mathbf{v}\right) \lambda^{k}=\exp \left[(x+2  \ii y+3 t) \lambda+4 t \lambda^{3}\right].\\
	\end{split}
	\end{equation}
	From the definition of $p_{k}^{[1]}(z)$ in Eq.(\ref{eq p}), if we set
	\begin{equation}\label{eq26}
	A=-3 t, \quad z=A^{-\frac{1}{3}}(x+3 t+2 \ii y),
	\end{equation}
	then the polynomial $p_{k}^{[1]}(z)$, $S_{k}(\textbf{x}^{+}(n)+\nu \textbf{s}-\mathbf{t})$ and $S_{k}\left(\mathbf{v}\right)$ satisfy
	\begin{equation}\label{eq78}
	S_{k}\left(\textbf{x}^{+}(n)+\nu \textbf{s}-\mathbf{t}\right)  \sim S_{k}\left(\mathbf{v}\right)=A^{\frac{k}{3}} p_{k}^{[1]}(z).
	\end{equation}

Based on the linear algebra, we can rewrite $\sigma_{n}(x-3t,y,t;\mathbf{A})$ as a $3N$ $\times$ $3N$ determinant:
	\begin{equation}\label{rewritten}
	\sum_{0 \leq \nu_{1}<\nu_{2}<\cdots<\nu_{N} \leq 2 N-1} \underset{1\leqslant i,j\leqslant N}{\det} \left[ \frac{1}{2^{\nu_{j}}} S_{2 i-1-\nu_{j}}\left(\textbf{x}^{+}(n)+\nu_{j} \textbf{s}-\mathbf{t}\right)\right] \times \underset{1\leqslant i,j\leqslant N}{\det} \left[\frac{1}{2^{\nu_{j}}} S_{2 i-1-\nu_{j}}\left(\textbf{x}^{-}(n)+\nu_{j} \textbf{s} -\mathbf{t}\right)\right].
	\end{equation}
By using the relationship (\ref{eq78}), we can reduce  the determinant involving $\textbf{x}^{+}(n)$ in  Eq.(\ref{rewritten}) as
		\begin{equation}\label{eq25}
		2^{-\delta} (-3 t)^{\frac{1}{3}\left(N^{2}-\delta\right)} \underset{1\leqslant i,j\leqslant N}{\det}\left[p_{2 i-1-\nu_{j}}^{[1]}(z)\right],
		\end{equation}
where $\delta=\nu_{1}+\nu_{2}+\cdots+\nu_{N}.$ It is obvious that the highest order term of $t$ can reach by choosing $\nu_j=j-1$, then we have
	\begin{equation}\label{eq26}
	\underset{1\leqslant i,j\leqslant N}{\det}\left[\frac{1}{2^{\nu_{j}}} S_{2 i-1-\nu_{j}}\left({\bf x}^{+}(n)+\nu_{j} {\bf s}-\mathbf{t}\right)\right]= 2^{-N(N-1) / 2}(-3 t)^{\frac{1}{3} \cdot\left(\frac{N(N+1)}{2}\right)} c_{N}^{-1} Q_{N}^{[1]}(z).
	\end{equation}
From the expansion  in Eq.(\ref{rewritten}), when $|t|\to\infty$, the leading order term of $\sigma_n(x-3t,y,t;\mathbf{A})$ is
	\begin{equation}\label{eq32}
	\left| 2^{-N(N-1)/2}(-3)^{\frac{N(N+1)}{6}}c_{N}^{-1}\right|^{2} t^{\frac{N(N+1)}{3}}\left|Q_{N}^{[1]}(z)\right|^{2}+\mathcal{O}(t^{\frac{N(N+1)-2}{3}}),\quad   |t|\to \infty.
	\end{equation}
	We can see that the leading terms of $\sigma_{n}(x-3t,y,t;\mathbf{A})$ are independent with $n$, therefore $u_{N}(x,y,t;\mathbf{A})=\left| \frac{\sigma _{1}\left( x-3t, y, t; \mathbf{A}\right) } {\sigma_{0}\left( x-3t, y, t; \mathbf{A}\right) }\right| ^{2}-1\to 0$ as $|t|\to \infty$. It completes the proof.

\setParDis
\textbf{\emph{Proof of Proposition \ref{property2}.}}
	\djy{When $t$ is large and $ |t|\gg |a_{2k+1}|^{\frac{3}{2k}}$, $(1 \leq k \leq N-1)$, in the neighborhood of $(x,y)=(x_{0},y_{0})$,} the coefficients about the highest term $t^{\frac{N(N+1)}{3}}$ of $\sigma_{n}(x-3t,y,t;\mathbf{A})$ will be almost to zero. Therefore we have to consider the second highest order term of $t$. Then we should make a more refined asymptotics for $S_{k}\left(\textbf{x}^{+}(n)+\nu \textbf{s}-\mathbf{t}\right)$:
	\setParDef
	\begin{equation}\label{eq34}
	\begin{split}
	S_{k}\left(\textbf{x}^{+}(n)+\nu \textbf{s}-\mathbf{t}\right) &=\epsilon^{-k} S_{k}\left(\left( x+3 t+2 \ii y+n\right)  \epsilon, \nu s_{2} \epsilon^{2}, \cdots\right) \\
	&=\epsilon^{-k} S_{k}\left(\left( x+3 t+2 \ii y+n\right) \epsilon, 0, \frac{\left(3^{3}-3\right) t}{3 !} \epsilon^{3},0,0,0, \cdots\right)\left[1+\mathcal{O}\left(\epsilon ^{2}\right)\right] \\
	&=S_{k}\left( \hat{\mathbf{v}}\right)\left[1+\mathcal{O}\left(\epsilon^{2}\right)\right],
	\end{split}
	\end{equation}
where  $\hat{\mathbf{v}}=(x+3 t+2 \ii y+n, 0,4 t, 0, 0,0,\ldots).$
	By setting $\hat{z}=A^{-\frac{1}{3}}(x+3 t+2 \ii y+n)$, then the Schur polynomial $S_{k}\left(\mathbf{x}^{+}(n)+\nu \mathbf{s}-\mathbf{t}\right)$ can be given with a similar formula as Eq.(\ref{eq26}). Especially, $p_{k}^{[1]}(\hat{z})$ and $S_{k}\left(\hat{\mathbf{v}}\right)$ have the following relation:
	\begin{equation}
	S_{k}\left(\textbf{x}^{+}(n)+\nu \textbf{s}-\mathbf{t}\right)=S_{k}\left(\hat{\mathbf{v}}\right)\left[1+\mathcal{O}\left(\epsilon^{2}\right)\right]=A^{\frac{k}{3}} p_{k}^{[1]}(\hat{z})\left[1+\mathcal{O}\left(\epsilon^{2}\right)\right].
	\end{equation}
	
	According to the equations Eq.(\ref{rewritten}), (\ref{eq25}), we know that the second highest order term is composed by two index choices of $\nu$: one is $\boldsymbol{\nu}=(0,1, \cdots, N-1)$ and the other one is $\boldsymbol{\nu}=(0,1, \cdots, N-2, N)$. Then the leading order terms should be the combination of these two factors. Firstly, if $\boldsymbol{\nu}=(0,1, \cdots, N-1)$, with a similar calculation as Eq.(\ref{eq25})$\sim$(\ref{eq32}), we can derive the determinant involving the $\textbf{x}^{+}(n)$ as
	\begin{equation} \label{eq31}
	\underset{1\leqslant i,j\leqslant N}{\det}\left[\frac{1}{2^{ \nu_{j}}} S_{2 i-1-\nu_{j}}\left({\bf x}^{+}(n)+\nu_{j} {\bf s}-\mathbf{t}\right)\right] =2^{-N(N-1)/2}(-3)^{\frac{N(N+1)}{6}}c_{N}^{-1} \cdot t^{ \frac{N(N+1)}{6}} Q_{N}^{[1]}(\hat{z})\left[1+\mathcal{O}\left(\epsilon^{2}\right)\right].
	\end{equation}
	As $Q_{N}^{[1]}(z_{0})=0$, we should expand $Q_{N}^{[1]}(\hat{z})$ at $\hat{z}=z_{0}$,
	\begin{equation}
	\begin{split}
	Q_{N}^{[1]}(\hat{z}) &=Q_{N}^{[1]}\left(z_{0}\right)+[Q_{N}^{[1]}]^{\prime}\left(z_{0}\right)\left(\hat{z}-z_{0}\right)+\mathcal{O}((\hat{z}-z_0)^2) \\
	&\left.=[Q_{N}^{[1]}]^{\prime}\left(z_{0}\right) A^{-\frac{1}{3}}\left[(x-x_{0}\right)+2 \ii \left(y-y_{0}\right)+n\right][1+\mathcal{O}(\epsilon)].
	\end{split}
	\end{equation}
Substitute the above equation into Eq.(\ref{eq31}), then the determinant involving $\mathbf{x}^{+}(n)$ changes into
	\begin{equation}\label{eq35}
 2^{-N(N-1)/2}(-3)^{\frac{N(N+1)-2}{6}}c_{N}^{-1}	\left[\left(x-x_{0}\right)+2 \ii\left(y-y_{0}\right)+n\right] t^{\frac{N(N+1)-2}{6}}\left[Q_{N}^{[1]}\right]^{\prime}\left(z_{0}\right)[1+\mathcal{O}(\epsilon)],
	\end{equation}
	Similarly, the determinant involving $\mathbf{x}^{-}(n)$ becomes
	\begin{equation}\label{eq36}
	2^{-N(N-1)/2}(-3)^{\frac{N(N+1)-2}{6}}c_{N}^{-1}\left[\left(x-x_{0}\right)-2 \ii\left(y-y_{0}\right)-n\right]  t^{\frac{N(N+1)-2}{6}}\left[Q_{N}^{[1]}\right]^{\prime}\left(z_{0}^{*}\right)[1+\mathcal{O}(\epsilon)].
	\end{equation}
	
	Now we consider the second part with $\boldsymbol{\nu}=(0,1, \cdots N-2 ,N)$, then the determinant involving $\textbf{x}^{+}(n)$ is
	\begin{equation}\label{eq37}
	\begin{split}
	& \underset{1\leqslant i,j\leqslant N}{\det}\left[\frac{1}{2^{v_{j}}} 		S_{2 i-1-\nu_{j}}\left(\textbf{x}^{+}(n)+\nu_{j} \textbf{s}-\mathbf{t}\right)\right] \\
	=& \underset{1\leqslant i,j\leqslant N}{\det}\left[\frac{1}{2^{0}} S_{2 i-1}\left(\textbf{x}^{+}-\mathbf{t}\right),  \cdots ,\frac{1}{2^{N-2}} S_{2 i-N+1}\left[\textbf{x}^{+}+\left( N-2\right)  \textbf{s}-\mathbf{t}\right], \frac{1}{2^{N}} S_{2 i-N-1}\left(\textbf{x}^{+}+N\textbf{s}-\mathbf{t}\right)\right]\\
	=& 2^{-\frac{N(N-1)}{2}-1} A^{\frac{N(N+1)-2}{6}} \underset{1\leqslant i,j\leqslant N}{\det} \left[p^{[1]}_{2 i-1}(\hat{z}), p^{[1]}_{2 i-2}(\hat{z}), \cdots p^{[1]}_{2 i-N+1}(\hat{z}), p^{[1]}_{2 i-N-1}(\hat{z})\right]\left[1+\mathcal{O}\left(\epsilon^{2}\right)\right].
	\end{split}
	\end{equation}
From the definition of $p^{[1]}_{k}(\hat{z})$, we have the relation $p^{[1]}_{2 i-(N+1)}(\hat{z})=[p_{2 i-N}^{[1]}]^{\prime}(\hat{z})$, which implies that the determinant in Eq.(\ref{eq37}) is $c_{N}^{-1}\left[Q^{[1]}_{N}\right]^{\prime}(\hat{z})$. Thus, Eq.(\ref{eq37}) will be simplified as
	\begin{equation}\label{eq38}
		2^{-\frac{N(N-1)}{2}}(-3)^{\frac{N(N+1)-2}{6}}c_{N}^{-1} t^{\frac{N(N+1)-2}{6}}\left[Q^{[1]}_{N}\right]^{\prime}(\hat{z})\left[1+\mathcal{O}\left(\epsilon^{2}\right)\right].
	\end{equation}
Furthermore, by expansion $\left[Q^{[1]}_{N}\right]^{\prime}(\hat{z})$ at $\hat{z}=z_{0}$, we have
	\begin{equation}
	\left[Q^{[1]}_{N}\right]^{\prime}(\hat{z})=\left[Q^{[1]}_{N}\right]^{\prime}\left(z_{0}\right)[1+\mathcal{O} \left( \epsilon)\right],
	\end{equation}
then the Eq.(\ref{eq38}) can be reduced into
	\begin{equation}\label{eq40}
 2^{-\frac{N(N-1)+2}{2}}(-3)^{\frac{N(N+1)-2}{6}}c_{N}^{-1} t^{\frac{N(N+1)-2}{6}}\left[Q^{[1]}_{N}\right]^{\prime}\left(z_{0}\right)[1+\mathcal{O}(\epsilon)].
	\end{equation}
Similarly, for the vector $\boldsymbol{\nu}=(0,1,\cdots,N-2, N)$, the determinant involving $\textbf{x}^{-}(n)$ is
	\begin{equation}\label{eq41}
 2^{-\frac{N(N-1)+2}{2}}(-3)^{\frac{N(N+1)-2}{6}}c_{N}^{-1} t^{\frac{N(N+1)-2}{6}}\left[Q^{[1]}_{N}\right]^{\prime}\left(z_{0}^{*}\right)[1+\mathcal{O}(\epsilon)].
	\end{equation}
Substitute the above Eq.(\ref{eq35}) (\ref{eq36}) (\ref{eq40}) (\ref{eq41}) to Eq.\eqref{rewritten}, we have
	\begin{equation}\label{eq46}
	\begin{split}
	\sigma_{n}(x-3t, y, t;\mathbf{A})=&\left|	2^{-\frac{N(N-1)}{2}}(-3)^{\frac{N(N+1)-2}{6}}c_{N}^{-1}\right| ^{2}\left|\left[Q^{[1]}_{N}\right]^{\prime}\left(z_{0}\right)\right|^{2} t^{\frac{N(N+1)-2}{3}}\\ &\cdot \left[\left(x-x_{0}\right)^{2}+4\left(y - y_{0}\right)^{2}-4 \ii n\left(y-y_{0}\right)-n^{2}+\frac{1}{4}\right] \left[1+\mathcal{O}(\epsilon)\right].
	\end{split}
	\end{equation}
 According to the properties of the root structure of $Q^{[1]}_{N}(z)$, we know $\left[Q^{[1]}_{N}\right]^{\prime}(z_{0}) \neq 0$, which means that the coefficient of the second highest term $t^{ \frac{N(N+1)-2}{3}}$ in Eq. (\ref{eq46}) never equals to zero. In this case, when $|t|$ is large, the asymptotic expression is
\begin{equation}
	u_{N}(x, y, t;\mathbf{A})=\left|\frac{\sigma_{1}(x-3t, y, t; \mathbf{A})}{\sigma_{0}(x-3t, y, t; \mathbf{A})}\right|^{2}-1=\frac{-32\left( x-x_{0}\right) ^{2}+128\left( y-y_{0}\right) ^{2}+8}{\left( 4\left(x- x_{0}\right) ^{2}+16\left(y-y_{0}\right)^{2}+1\right) ^{2}}+\mathcal{O}(\epsilon).
\end{equation}
	Clearly, the $N$-th order lump solution approaches to the first-order lump solution $u_{1}(x-(\hat{x}_{0}+3t),y-\hat{y}_{0},t)$ as $t$ is large with the error $\mathcal{O}(|t|^{-\frac{1}{3}})$. It completes the proof.

       	\setParDis
	 \textbf{\emph{Proof of Proposition \ref{property3}.}}  In  Ref.\cite{yang2021rogue}, the authors had studied the rogue wave pattern near the origin for NLS equation. Similarly, we can analyze the asymptotics for the high order lump solutions. Compared to the analysis for rogue waves, we only need to make a modified definition to the vectors $\textbf{x}^{\pm}$ and $\textbf{y}^{\pm}$. In this paper, we define $\textbf{y}^{\pm}$ as
	 	\setParDef
	\begin{equation}
	\textbf{x}^{+}-\mathbf{t}=\textbf{y}^{+}+(0,0,4t,0,\dots),\quad \textbf{x}^{-}-\mathbf{t}=\textbf{y}^{-}+(0,0,4t,0,\dots).
	\end{equation}
	Following the method in \cite{ohta2012general}, $\sigma_{n}$ can be written as a $3N\times 3N$ determinant form
	\begin{equation}
	\begin{vmatrix}
	\mathrm{O}_{N\times N}&\Omega_{N\times 2N}^{(n)}\\
	-\Psi_{2N\times N}^{(n)}&\mathrm{I}_{2N\times 2N}
	\end{vmatrix},
	\end{equation}
	where $\Omega_{i,j}=\frac{1}{2^{j-1}}S_{2i-j}[\textbf{x}^{+}(n)+(j-1)\textbf{s}-\mathbf{t}]$ and $\Psi_{i,j}=\frac{1}{2^{i-1}}S_{2j-i}[\textbf{x}^{-}(n)+(i-1)\textbf{s}-\mathbf{t}]$. By a simple calculation to the determinant $\sigma_{n}$, we get the asymptotics as
	\begin{equation}
	\sigma_{n}=\beta\left|t\right|^{k^{2}(2 m+1)+k\left(2 N_{0}+1\right)}\left|\begin{array}{cc}
	\mathbf{O}_{N \times N} & \widetilde{\Omega}_{N \times 2 N} \\
	-\widetilde{\Psi}_{2 N \times N} & \mathbf{I}_{2 N \times 2 N}
	\end{array}\right|\left[1+\mathcal{O}\left(\left|t\right|^{-1}\right)\right],
	\end{equation}
	where $\widetilde{\Omega},\widetilde{\Psi},\beta$ are defined as Ref.\cite{yang2021rogue}. With a similar analysis, in the neighborhood of $(x,y)=(-3t, 0)$, we have the following asymptotics:	
	\begin{equation}
	u_{N}\left(x, y,t;\mathbf{A} \right)=\left|\frac{\sigma_{1}(x-3t, y, t; \mathbf{A})}{\sigma_{0}(x-3t, y, t; \mathbf{A})} \right|^{2}-1=u_{N_{0}^{[1]}}\left(x, y,t\right)\left[1+\mathcal{O}\left(\left|t\right|^{-1}\right)\right], \quad |t| \to \infty.
	\end{equation}
According to Eq.(\ref{N02}), we know that $N_{0}^{[1]}$ can only be $0$ or $1$. It completes the proof.
	
	\setParDis
\textbf{\emph{Proof of Proposition \ref{property4}.}}
	 When an internal parameter $a_{2m+1}$ is large enough: \djy{$|a_{2m+1}|\gg \frac{2^{2m}}{2m+1}, (1<m\leq N-1)$ and other parameters satisfy $a_{2k+1}=o(|a_{2m+1}|^{\frac{2k}{2m+1}}),(k\neq m)$,}  we need to compare the effect of $a_{2m+1}$ and $t$. If $t$ is the dominant term and satisfies the condition $|t|\gg |a_{2m+1}|^{\frac{3}{2m}}$,  that is $a_{2m+1}=o\left(t^{\frac{2m}{3}}\right)=o\left(\epsilon ^{-2m}\right)$, we have $\left( x_{2m+1}^{\pm}-\frac{3t}{\left( 2m+1\right) !}\right) \epsilon^{2m+1}=o(\epsilon)$ for $m \neq  1$, then the proposition \ref{property1}-\ref{property3} are still available. Therefore, the high order lump solutions can be separated into $N^{[1]}_{p}$ first-order lumps  and theorem \ref{theo2} is still valid. When $|t|\ll |a_{2m+1}|^{\frac{2}{2m+1}}$, that is $t={o}\left(|a_{2m+1}|^{\frac{2}{2m+1}}\right)$, we need to analyze the asymptotics of lump solutions with respect to $a_{2m+1}$. This case is similar to the asymptotics of rogue waves for NLS equation.
	\setParDef
	
	 In Ref. \cite{yang2021rogue}, the authors discussed the asymptotics of rogue waves  about the multiple internal parameters for NLS equation.
	Similar to this result,
	when $(x,y)$ is far away from $(-3t,0)$ with $ \sqrt{(x+3 t)^{2}+4y^{2}}=\mathcal{O} \left(\left|a_{2 m+1}\right|^{\frac{1}{2 m+1}}\right)$, we set $\hat{\epsilon}=|a_{2m+1}|^{-1/(2m+1)}$, which indicates $t=o(\hat{\epsilon}^{-2})$, then the Schur polynomials in theorem \ref{theo2} change into
	\begin{equation}
	\begin{split}
	S_{k}\left(\textbf{x}^{+}(n)+\nu \textbf{s}-\mathbf{t}\right) &=\hat{\epsilon}^{-k} S_{k}\left((x+3 t+2 \ii y+n) \hat{\epsilon}, \nu s_{2} \hat{\epsilon}^{2}, \cdots\right) \\
	&=\hat{\epsilon}^{-k} S_{k}\left((x+3 t+2 \ii y+n) \hat{\epsilon}, 0,\cdots,0,1,0, \cdots\right)\left[1+o\left(\hat{\epsilon}\right)\right] \\
	&=S_{k}(x+3 t+2 \ii y+n, 0,\cdots,0,a_{2m+1}, 0, \cdots )\left[1+o\left(\hat{\epsilon}\right)\right].
	\end{split}
	\end{equation}
	%This relation is the counterpart of Eq.(\ref{eq34}) during the proof of proposition \ref{property2}.
Similarly, when $t$ is large, we can get the highest order about $a_{2m+1}$ as
	\begin{equation}
	2^{-N\left( N-1\right) } c_{N}^{-2}\left(-\frac{2 m+1}{2^{2 m}}\right)^{\frac{N(N+1)}{2 m+1}}\left|a_{2 m+1}\right|^{\frac{N(N+1)}{2 m+1}}\left|Q_{N}^{[m]}\left(z\right)\right|^{2},
	\end{equation}
	where
	\begin{equation}
		z=\hat{A}^{-\frac{1}{2m+1}}(x+3t+2\ii y),\quad \hat{A}=-\frac{2m+1}{2^{2m}}a_{2m+1},
	\end{equation}
	and $c_{N}^{-1}$ and $Q_{N}^{[m]}\left(z\right)$ are defined as Eq.(\ref{eq p}). Thus when $Q_{N}^{[m]}\left(z\right)\neq 0$, we have $u_{N}(x,y,t;\mathbf{A}) \to 0$. But when $Q_{N}^{[m]}\left(z\right)= 0$, the highest order term will be zero, we can consider the asymptotics in the neighbourhood of the special point $(x,y,t)=(x_{0}^{[m]},y_{0}^{[m]},t)$ with the definition
	\begin{equation}
	x_{0}^{[m]}+3t+2\ii  y_{0}^{[m]}=\hat{A}^{\frac{1}{2m+1}}z_{0}^{[m]},
	\end{equation}
	where $z_{0}^{[m]}$ is the root of Yablonskii-Vorob'ev polynomials $ Q_{N}^{[m]}\left(z\right)$, which has $N_{p}^{[m]}$ simple nonzero roots.
	Similar to the proof of proposition \ref{property2}, if $(x,y)$ satisfies
$\left[\left(x-x_{0}^{(m)}\right)^{2}+4\left(y-y_{0}^{(m)}\right)^{2}\right]^{1 / 2}=\mathcal{O}(1),$we have
	\begin{equation}
	u_{N}(x, y, t;\mathbf{A})=\left|\frac{\sigma_{1}(x-3t, y, t; \mathbf{A})}{\sigma_{0}(x-3t, y, t; \mathbf{A})}\right|^{2}-1
=\frac{-32\left( x-x_{0}^{[m]}\right) ^{2}+128\left( y-y_{0}^{[m]}\right) ^{2}+8}{\left(4\left(x- x_{0}^{[m]}\right)^{2}+16\left(y-y_{0}^{[m]}\right)^{2}+1\right)^{2}}+\mathcal{O}(\hat{\epsilon}).
	\end{equation}
Then we can get a conclusion that, when $a_{2m+1}$ is large enough,  in the neighborhood of $(x,y)=(-3t,0)$, the corresponding calculation is similar to the solution of NLS equation near the  origin with one parameter large. Bo Yang et. al. had studied this case for NLS equation in theorem 4 in Ref\cite{yang2021rogue}. Hence,  in the neighborhood of $(x,y)=(-3t,0)$, the lump solution will decay to a ${N}_{0}^{[m]}$-th order lump solution. It completes the proof.

%\bibliographystyle{unsrt}
%\bibliography{ref}
\end{document}